\documentclass[aps,prd,twocolumn,amsmath,superscriptaddress,amssymb,showpacs,floatfix,nofootinbib, superscriptaddress, longbibliography]{revtex4-1}

\usepackage{graphicx}
\usepackage{dcolumn}
\usepackage{xcolor}
\usepackage{bm}
\usepackage{natbib}
\usepackage{calligra}
\newfont{\rsfsten}{rsfs10 scaled 1200}
\newfont{\rsfsseven}{rsfs10 scaled 1200}
\newfont{\rsfsfive}{rsfs10 scaled 1200}
\usepackage{hyperref}
\usepackage[utf8]{inputenc}

\newcommand{\aap}{{Astron.~Astrophys.}}

\newcommand{\aj}{{Astron.~J.}}

\newcommand{\mnras}{{Mon.~Not.~R.~Astron.~Soc.}}

\newcommand{\BHprime}{BH$^{\prime} \,$}

\begin{document}
	
\hspace{13cm}
\space{0.6cm}
	


\title{Black Holes merging with Low Mass Gap Objects inside Globular Clusters}
	
\author{Konstantinos Kritos}
\email{ge16004@central.ntua.gr}
\affiliation{Physics Division, National Technical University of Athens, Zografou, Athens, 15780, Greece}
\author{Ilias Cholis}
\email{cholis@oakland.edu, ORCID: orcid.org/0000-0002-3805-6478}
\affiliation{Department of Physics, Oakland University, Rochester, Michigan, 48309, USA}
	
\date{\today}

\begin{abstract}

Recently, the LIGO-Virgo collaborations have reported the coalescence of a binary involving a 
black hole and a low-mass gap object (LMGO) with mass in the range $\sim2.5-5M_\odot$. 
Such detections, challenge our understanding of the black hole and neutron star mass spectrum,
as well as how such binaries evolve especially if isolated. 
In this work we study the dynamical formation of compact object pairs, via multiple binary-single 
exchanges that occur at the cores of globular clusters. We start with a population of binary 
star systems, which interact with single compact objects as first generation black holes and LMGOs.
We evaluate the rate of exchange interactions leading to the formation of compact object binaries. 
Our calculations include all possible types of binary-single exchange interactions and also the 
interactions of individual stars with compact object binaries that can evolve their orbital properties, 
leading to their eventual merger. We perform our calculations for the full range of the observed 
Milky Way globular cluster environments.
We find that the exchanges are efficient in forming hard compact object binaries at the cores of dense 
astrophysical stellar environments. Furthermore, if the population size of the LMGOs is related to 
that of neutron stars, the inferred merger rate density of black hole-LMGO binaries inside globular 
clusters in the local Universe is estimated to be about $0.1 \, \text{Gpc}^{-3}\text{yr}^{-1}$.

\end{abstract}
	
\maketitle
	
\section{Introduction}
	
\vskip 0.05in

Globular clusters are complex environments where single and binary stars, neutron stars (NSs) and 
black holes (BHs), coexist and can have strong dynamical interactions. Those interactions can lead 
to the creation of compact object-star or compact object-compact object binaries. In the Milky Way we 
have observed so far more than 160 such systems for many of which we have well-measured total 
mass and density profiles \cite{2010arXiv1012.3224H, harris1996}. Moreover, in some of these 
environments the compact objects will have multiple interactions as has been shown in 
\cite{Sesana:2006xw, Sesana:2015haa, Rodriguez:2016kxx, Gupta:2019nwj, Fragione:2020wac, 
Fragione:2020nib, Samsing:2020qqd, Kritos:2020fjw}. Thus, we expect different combinations of 
binaries between the stars and the compact objects and the compact objects themselves, which may 
even lead to runaway mergers \cite{Miller:2001ez, PortegiesZwart:2002iks, AtakanGurkan:2003hm, 
OLeary:2005vqo, Kovetz:2018vly, Antonini:2018auk, Flitter:2020bky, Baibhav:2020xdf, Kritos:2020wcl}.

One exciting possibility is that among the many compact objects inhabiting globular clusters, are low 
mass gap objects (LMGOs). These are objects with mass larger than the expected NS upper mass 
limit of approximately 2.5 $M_{\odot}$ (see \cite{Ozel:2016oaf} for a relevant review), and lower 
than the empirical lower stellar-mass BHs from core collapses of 5 $M_{\odot}$ \cite{1998ApJ...499..367B, 
Fishbach:2017zga, Wysocki:2018mpo} (see however \cite{Thompson:2018ycv}). In fact, recently 
the Virgo and the LIGO Scientific collaborations reported the detection of a merger event (GW190814) 
between a $23.2^{+1.1}_{-1.0} \; M_\odot$ BH and a $2.59^{+0.08}_{-0.09} \; M_\odot$ LMGO 
\cite{Abbott:2020khf}. This discovery is even more intriguing due to its unusual high mass ratio. 
Its possible explanations include the discovery of the least massive BH \cite{Tews:2020ylw, 
Takhistov:2020vxs}, the most massive NS  \cite{Zhang:2020zsc,Most:2020bba,Tan:2020ics,
Tsokaros:2020hli,Biswas:2020xna,Lim:2020zvx,Godzieba:2020tjn,Dexheimer:2020rlp} ever observed,
an  exotic  compact object  \cite{Bombaci:2020vgw,Cao:2020zxi,Moffat:2020jic,Astashenok:2020qds} 
or a primordial black hole \cite{Clesse:2020ghq,Jedamzik:2020omx,Vattis:2020iuz}. 
Additionally, such BH-LMGO pairs may be formed inside dense clusters through complex triple and 
hierarchical quadruple systems \cite{Safarzadeh:2019qkk,Lu:2020gfh,Sedda:2020wzl,Rastello:2020sru,
Sedda:2021oov}, or as isolated pairs \cite{Zevin:2020gma}, or be the result of mass accretion of a NS 
from the supernova envelope of its progenitor in an isolated compact binary \cite{Safarzadeh:2020ntc}. 
Finally, the peculiar derived compact objects mass-values may just be due to GW190814 being a lensed 
merger event \cite{Broadhurst:2020cvm}.
	
Over the past few years the LIGO-Virgo collaboration have detected double neutron star mergers 
\cite{TheLIGOScientific:2017qsa,Abbott:2020uma}, with the merged product being in the mass gap interval. 
A fraction of objects in the mass range of LMGOs certainly originated from the coalescence of two neutron 
stars with the product having a mass around 3$M_\odot$. While this may not be the only mechanism to create 
LMGOs (see for instance \cite{Yang:2020xyi}), in this work we are not concerned with the nature of the LMGOs. 
Instead, we take their number solely coming from double neutron star mergers. This is a conservative 
choice which also uses the fact that the 2.59$M_\odot$ object lies in the mass range of double neutron 
star mergers \cite{Gupta:2019nwj}. 
	
We implement a simple model to evaluate the merger rate of BH-LMGO mergers formed in Milky Way 
globular cluster-type environments. We start with star-star binaries and through successive binary-single 
exchanges end in creating BH-BH, BH-LMGO and LMGO-LMGO binaries. We monitor the different populations 
of single compact objects i.e. BHs and LMGOs, as well as the different combinations of star/LMGO/BH binary 
populations.  We rely on a qualitative description of the mass function for the populations we consider, and 
take monochromatic mass spectra for the stars, the LMGOs and the BHs; but also include the presence of 
second generation black holes with mass nearly double that of the first generation ones.  We demonstrate that BH-LMGO 
pairs can be produced and coalesce in a dynamical environment, where successive exchange events and 
hardening interactions with third less massive objects take place.   

This paper is organized as follows. In section~\ref{secMethod} we describe the setup to calculate the formation 
of compact binaries via exchanges inside stellar clusters and also how their orbital properties evolve. Then, in 
section \ref{secResults} we present and discuss our findings on the time evolution of binary populations and 
their merger rates. Our work provides a realistic estimate of the minimum contribution globular cluster 
environments have on the BH-LMGO merger rates.  Finally, we give our conclusions in section~\ref{secConclusions}. 

\section{Methodolody and Assumptions}
\label{secMethod}
\begin{figure*}
\includegraphics[width=0.95\textwidth]{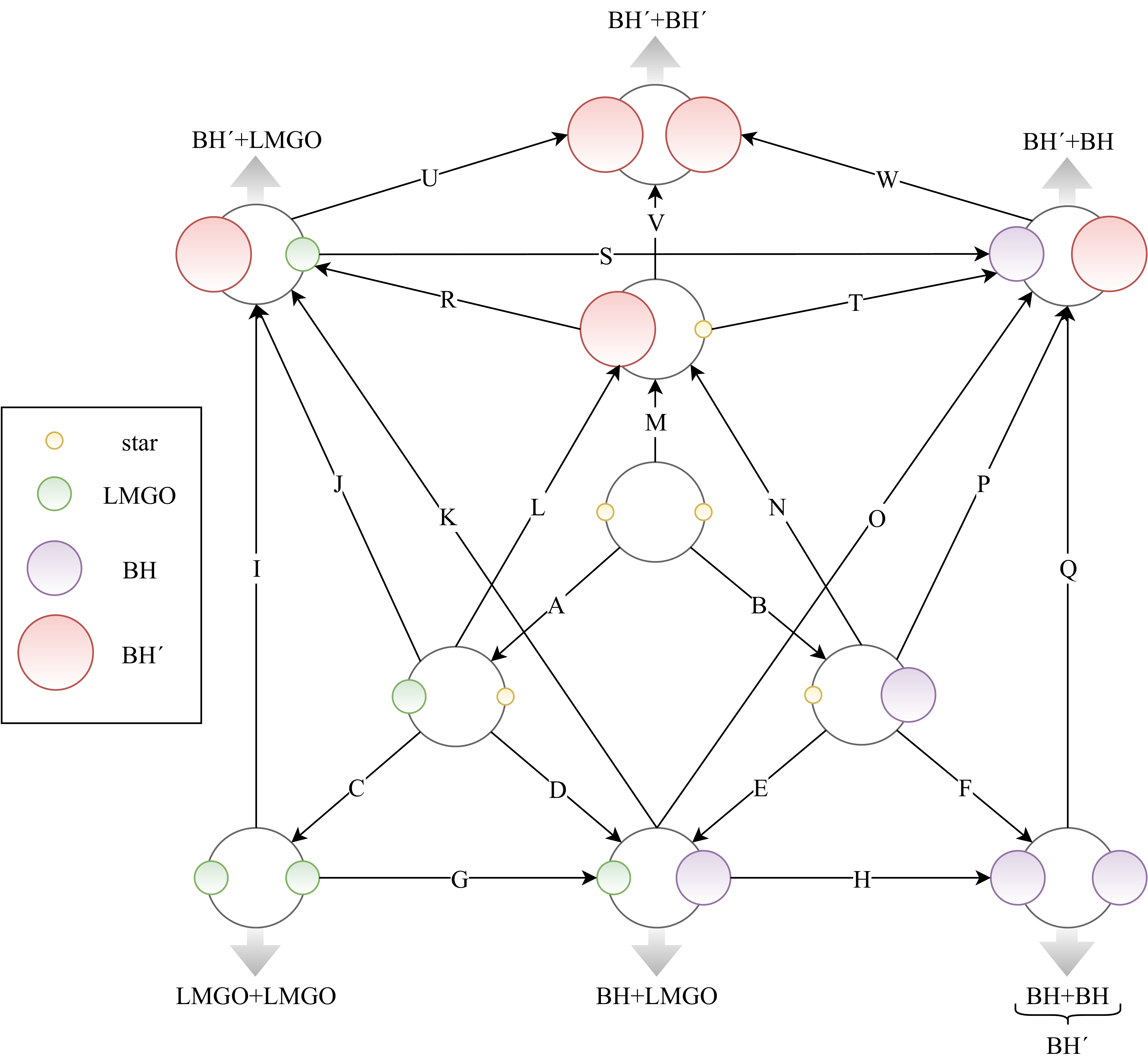}
\caption{A representation of the processes that we take into account for the formation of compact binaries 
inside globular clusters. Each narrow arrow implies a binary-single exchange and the corresponding rate 
contributes to the right hand side of the differential system in Eq.~\ref{differentialSystem}. Thick arrows 
represent the merger of binaries into more massive compact objects. In this work we take as reference 
values $m_\text{star}=1 M_\odot$, $m_\text{LMGO}=3 M_\odot$, $m_\text{BH}=10 M_\odot$ and 
$m_{\textrm{\BHprime}} = 19 M_{\odot}$.}
\label{diagram}
\end{figure*}

We assume that initially all LMGOs and all first generation black holes which we denote from here as ``BH'' start 
as single objects in the history of a given cluster. Only stars initially are in binaries. In this work compact object binaries 
are assembled through multiple exchanges from the smaller mass star-star pairs. We focus 
on the dynamical interactions between binaries and individual objects and take an agnostic approach on 
the origin of LMGOs, just assuming their number is set by the NS that remain in the clusters after the natal
kicks. Binary stars can undergo exchange events with individual LMGOs or BHs leading to the creation of
LMGO-star or BH-star binaries respectively. In Fig.~\ref{diagram} we depict those exchange interactions as channels
``A'' and ``B''. Those binaries can then undergo further exchanges through channels ``C'', ``D'', ``E'' and ``F'' 
and create LMGO-LMGO, BH-LMGO or BH-BH binaries. In fact when the individual BHs come at close proximity 
to either the LMGO-LMGO or the BH-LMGO binaries further exchanges may take place leading to channels ``G'' 
and ``H''.  

All exchange channels are depicted in Fig.~\ref{diagram} by thin arrows and are identified by a letter.  There is a flow 
of energy transferred from the pairs involving smaller mass objects into binaries involving 
BHs and LMGOs which then consequently merge via 3rd-body hardening evolution. The merger evens are depicted 
as thick arrows and can lead to black holes. Those black holes can then be kicked out of the cluster due to a gravitational wave kick
or remain in it and enter the population of compact objects available to dynamically interact. Due to what we will show to be a 
small number of LMGO-LMGO pairs and BH-LMGO pairs we ignore those objects. However, the BH-BH binaries merge into 
second generation black holes with mass approximately double the original first generation one. Those second generation
black holes we denote as ``\BHprime ''. The \BHprime objects that will remain in the clusters will then participate in exchange 
interactions with all type of existing binaries in the cluster creating also \BHprime-star, \BHprime-LMGO and \BHprime-BH binaries,
through the exchange channels ``I'', through``Q''. 
Those types of binaries are also connected through exchange interactions (channels ``S'', ``R'' and ``T'').
Finally, ``\BHprime-\BHprime '' binaries can be created through exchange interactions of channels ``U'', ``V'' and ``W''.   

In the following we describe all specific assumptions on the creation and evolution of these binaries.

\subsection{Hard binaries}
	
A binary $1-2$ with mass components $m_1$ and $m_2$ is defined as hard, when its semimajor axis (SMA) is 
smaller than $a_H$. We will refer to $a_H$ as hardness SMA. This also defines the semimajor axis value 
below which a pair hardens at a constant rate via 3rd-body encounters with single stars of mass $m_*$. 
Assuming that $m_1\ge m_2\gg m_*$, this hardness SMA is given by \cite{Quinlan:1996vp,Sesana:2006xw},
\begin{equation}
a_H\equiv{Gm_2\over4\sigma_*^2}\approx22.3\times\left({m_2\over10M_\odot}\right)\left({\sigma_*\over10\text{km/s}}\right)^{-2}\text{AU}.
\label{eq:aH}	
\end{equation}
$\sigma_{*}$, is the velocity dispersion of stars (see Appendix~\ref{AppVelos} for further details).
	
After a single 3rd-body encounter the orbital SMA 
and eccentricity change on average to, 
\begin{eqnarray}
\label{eq:SMAevolHard}
a'&=a\left[1+{H\over2\pi}{m_*\over m_1+m_2}\right]^{-1}, \\
e'&=e+K\left[1+{2\pi\over H}{m_1+m_2\over m_*}\right]^{-1} ,
\label{eq:ECCevolHard}
\end{eqnarray}
where $H$ and $K$ are two parameters determined numerically.  Past surveys 
\cite{Quinlan:1996vp,Sesana:2006xw} have showed that $H$ lies in the interval [15,20] and $K$ is in [0,0.2]. 
We will use as reference the values of $H=20$ and $K=0.05$ for these parameters and will describe the 
impact of alternative choices which we find change only marginally our results.

Tight binaries harden until their SMA passes a critical value which we denote by $a_\text{GW}$. That SMA
corresponds to the point where GW emission starts to dominate over the 3rd-body hardening process. 
Given that $\rho_*$ is the mass density of stars and $\sigma_{*,1-2}$ is the rms velocity of stars relative 
to the $1-2$ pair (see Appendix~\ref{AppVelos}), then this critical SMA is approximately equal to, 
\cite{Sesana:2015haa,Quinlan:1996vp},
\begin{eqnarray}
a_\text{GW}&\simeq&0.05\text{AU}\times\left({\sigma_{*,1-2}\over10\text{km/s}}\right)^{1/5}
\left({\rho_*\over10^5M_\odot/\text{pc}^3}\right)^{-1/5} \nonumber \\
&\times&\left({H\over20}\right)^{-1/5}\left({m_1\over10M_\odot}\right)^{1/5}\left({m_2\over10M_\odot}\right)^{1/5}
\nonumber \\
&\times& \left({m_1+m_2\over20M_\odot}\right)^{1/5} (1-e_\text{GW}^2)^{-7/10} \nonumber \\
&\times&\left(1+{73\over24}e_\text{GW}^2+{37\over96}e_\text{GW}^4\right)^{1/5},
\label{eq:SMAevol}
\end{eqnarray}
where $e_\text{GW}$ is the eccentricity of the pair at the point when GW emission starts to dominate. 
After the binary enters this GW domination phase, the timescale required for it to merge solely due 
to the emission of GWs is given by \cite{1964PhRv..136.1224P},
\begin{eqnarray}
T_\text{GW}&\approx&1.5\times10^{11}\text{yr}\left({m_1\over10M_\odot}\right)^{-1}\left({m_2\over3M_\odot}
\right)^{-1} \nonumber \\
&\times& \left({m_1+m_2\over13M_\odot}\right)^{-1} \left({a_\text{GW}\over0.1\text{AU}}\right)^{4}(1-e_\text{GW}^2)^{7/2}.
\label{eq:Tgw}
\end{eqnarray}
In particular, if for instance $e_\text{GW}=0.9$, then the above timescale is about 440Myr. For a 
numerical description of the 3rd-body and gravitational wave emission inside realistic globular 
cluster environments one can see \cite{Kritos:2020fjw, Samsing:2020qqd}. 

In this work for all binaries 
made out of two compact objects (i.e. LMGO-LMGO, BH-LMGO, \BHprime-LMGO, BH-BH, \BHprime-BH and 
\BHprime-\BHprime) we include the effect of 3rd-body hardening interactions and evolve their SMA until they 
reach their respective value of $a_\text{GW}$; calculated by Eq.~\ref{eq:SMAevol} for the given 
binary masses $m_{1}$, $m_{2}$ and globular cluster environment ($\sigma_{*}$, $\rho_{*}$). Once 
the binaries reach their respective $a_\textrm{GW}$, they will merge within $T_\text{GW}$ 
calculated from Eq.~\ref{eq:Tgw}.

\subsection{Number of compact objects in stellar clusters}
	
Ref.~\cite{Breen:2013vla} has showed that once a system of stellar mass black holes is created in the 
core of a globular cluster it remains for about 10 times the initial dynamical relaxation time. That 
relaxation time for the larger clusters relevant in this work, is more than a Gyr. Also, 
Refs.~\cite{2015ApJ...800....9M, 2019ApJ...871...38K} have shown that large numbers of black holes are 
retained inside globular clusters for a wide range of their properties. Recently Ref.~\cite{2021A&A...646A..63V} 
analyzed observational data from NGC 6397 supporting the presence of a core of compact objects at the 
center of that globular cluster. We choose for 
simplicity to take the volume that the stellar-mass black holes occupy $V_{\textrm{BH}}$, to be 
constant with time, and evolve our systems up to the Hubble time.
	
The number of first generation BHs in a given cluster is estimated from the initial stellar mass function 
and the natal kicks of the BHs. We take that only stars with mass larger that $25M_\odot$ have collapsed into a BH.
Based on the Kroupa stellar mass function \cite{Kroupa:2002ky}, $\approx10\%$ of a cluster's mass $M_\text{cl}$ 
has been in these stars. 
We assume that approximately $1/3$ of the progenitor's mass collapses to the BH remnant \cite{Mirabel:2016msh}. 
Thus the total mass in first generation BHs that have been produced in the history of the cluster is about $3\%$ of 
the cluster's mass, $f_\text{BH}\approx3\%$. 
Of them only a fraction $f_\text{ret}$ has  been retained by the gravitational potential 
well of the cluster while the rest escaped due to their natal kick. For each cluster we evaluate $f_\text{ret}$; 
on average we find it to be $10\%$.
As we take all first generation BHs to be of mass $m_\text{BH}$, the number of retained BHs in a cluster is,
\begin{eqnarray}
N_\text{BH}&=&f_\text{ret}f_\text{BH}{M_\text{cl}\over m_\text{BH}}\\&\approx& 30\times\left({f_\text{ret}\over10\%}\right)\left({f_\text{BH}\over3\%}\right)\left({M_\text{cl}\over10^5M_\odot}\right)\left({m_\text{BH}\over10M_\odot}\right)^{-1}\nonumber.
\end{eqnarray}
	
In a similar manner, we calculate the number of NSs, where for their retention fraction we have found it to 
be typically $f_\textrm{ret}^{\textrm{NS}} \simeq 2 \%$, relying on~\cite{Hansen:1997zw} for their observed 
natal kicks distribution. As for every cluster the $f_\textrm{ret}$ depends on the specific escape velocity, the 
more massive clusters retain a larger fraction of a larger number of such objects created in the first place.
The fraction of the cluster's mass corresponding to NSs is also $f_\text{NS}\approx3\%$, assuming that on 
average 1/3 of the progenitor's mass collapses to a NS. 
 
Regarding the number of produced LMGOs, we assume that they originate from NS-NS mergers and 
that all NSs that didn't escape the cluster merged into LMGOs. Thus the number of LMGOs  is simply half the 
number of the retained NSs, i.e. $N_{\text{NSNS}\to\text{LMGO}}={N_\text{NS}\over2}$ 
We also assume 
that those mergers happen very fast in the history of the cluster and essentially provide an initial population of 
LMGOs. However, we also study the case where the NS-NS binaries did not merge early in the history of the 
clusters, but instead evolve as a separate population of binaries. In that case those pairs progressively
give LMGOs which once produced are added to the population of single LMGOs and then are allowed to form 
binaries following the scheme of Fig.~\ref{diagram}.
The LMGOs from NS-NS mergers provide a candidate for these objects. The actual population 
size of LMGOs could be even larger if more exotic species of LMGOs are present as e.g. \cite{Bombaci:2020vgw, 
Roupas:2020nua, Kritos:2020wcl, Rather:2020lsg, Zhang:2020dfi}. In Appendix~\ref{AppParameters} we provide
results for different options setting the population size of LMGOs inside the cluster.

\subsection{Description of the model and initial conditions}

The population of BH-LMGO pairs can be assembled through the exchanges designated by the arrow 
sequences A+D or B+E or A+C+G in Fig.~\ref{diagram}. All pairs we consider are hard and do not get disrupted 
by stars \cite{1975MNRAS.173..729H}.  We take that all binaries when first assembled have the largest possible 
SMA and evolve it with time. For the exchange events the new binary starts with a SMA that has the same 
value as that of the progenitor binary. This are conservative choices as they lead to the smaller number of 
coalescence  events.
	
Regarding the SMA of the binaries, we consider a discretization of the SMA interval into bins and evolve each 
population in time. 
As a hard binary interacts with the surrounding stars, its SMA value becomes 
smaller by a factor of $a'/a$ evaluated from Eq.~\ref{eq:SMAevolHard} 
\footnote{We take a logarithmically spaced binning with 100 slots in total. Every binary, 
for example a BH-LMGO will harden due 3rd-body interactions. For instance after its i-th such interaction 
$a_\text{BH-LMGO}(i)=a_\text{BH-LMGO}(i-1)\cdot f_\text{H,BH-LMGO}$; where $a_\text{BH-LMGO}(1)
=a_\text{H,BH-LMGO}$ from Eq.~\ref{eq:aH} and $f_\text{H,BH-LMGO}$ 
is evaluated from Eq.~\ref{eq:SMAevolHard} for $m_{*} = 1 M_{\odot}$, $m_{1} = 10 M_{\odot}$, 
$m_{2} = 3 M_{\odot}$. As $a_\text{BH-LMGO}$ is reduced the BH-LMGO binary will ``jump'' between these slots.}.
That``jump'' is performed following a rate of interaction between the binary and any close-by star.
The velocity averaged interaction rate density at which a population of hard binaries $1-2$ with 
mass components $m_1$ and $m_2$ and number density $n_{1-2}$ interact with single objects 
$3$ with mass $m_3$ and number density $n_3$ is given by \cite{Colpi:2003wb},
\begin{equation}
\gamma_\text{int}=n_{1-2}n_3{2\sqrt{6\pi}G(m_1+m_2+m_3)\over \sigma_{3,1-2}}a.
\label{interRate}
\end{equation}
\vskip 0.05in
Here, $a$ is the SMA of the $1-2$ system, $G$ is the universal gravitational constant and 
$\sigma_{3,1-2}=\sqrt{\langle v_{3,1-2}^2\rangle}$ is the three dimensional velocity 
dispersion between the binary and the single object at infinity (see Appendix~\ref{AppVelos}). 
We note that, Eq.~\ref{interRate} is to be considered only at the gravitational focusing
 domination regime, i.e. holds valid only for strong interactions of stars with hard binaries.
The total interaction rate $\Gamma_{\textrm{int}}$ between a binary system $1-2$ with a
3rd body is just, 
\begin{equation}
\Gamma_{\textrm{int}} = \int_{0}^{V_{\textrm{BH}}} \gamma_{\textrm{int}} dV.
\label{eq:3rdBodyGamma}
\end{equation}

The set of differential equation describing the evolution of the number of single compact objects
and their binaries is,	
\begin{widetext}
	\begin{subequations}
		\begin{align}
		        \label{eq:LMGO}
			\dot{N}_\text{LMGO}&=-\Gamma_\text{A}-\Gamma_\text{C}-\Gamma_\text{E}+\Gamma_{\textrm{G}}+\Gamma_{\textrm{H}}+\Gamma_{\textrm{I}} +\Gamma_{\textrm{L}}+\Gamma_{\textrm{O}} -\Gamma_{\textrm{R}} +\Gamma_{\textrm{S}} + \Gamma_{\textrm{U}},\\
			\dot{N}_\text{BH}&=-\Gamma_\text{B}-\Gamma_\text{D}-\Gamma_\text{F} -\Gamma_\text{G} -\Gamma_\text{H}+\Gamma_\text{K}+\Gamma_\text{N}+\Gamma_\text{Q}-\Gamma_\text{S}-\Gamma_\text{T}+\Gamma_\text{W},\\
			\dot{N}_\text{\BHprime}&=+F_\text{gw}\Gamma_\text{BH+BH}-\Gamma_\text{I}-\Gamma_\text{J}-\Gamma_\text{K}-\Gamma_\text{L}-\Gamma_\text{M}-\Gamma_\text{N}-\Gamma_\text{O}-\Gamma_\text{P}-\Gamma_\text{Q}-\Gamma_\text{U}-\Gamma_\text{V}-\Gamma_\text{W},\\
			&---------------------- \nonumber
			\\
			\label{eq:LMGOSTAR}
			\dot{N}_\text{LMGO-STAR}&=+\Gamma_\text{A}-\Gamma_\text{C}-\Gamma_\text{D}-\Gamma_\text{J}-\Gamma_\text{L},\\
			\dot{N}_\text{BH-STAR}&=+\Gamma_\text{B}-\Gamma_\text{E}-\Gamma_\text{F}-\Gamma_\text{N}-\Gamma_\text{P},\\
			\label{eq:BHPSTAR}
			\dot{N}_\text{\BHprime-STAR}&=+\Gamma_\text{L}+\Gamma_\text{M}+\Gamma_\text{N}-\Gamma_\text{R}-\Gamma_\text{V}-\Gamma_\text{T},\\
			&---------------------- \nonumber
			\\
			\label{eq:BHLMGO}
			\dot{N}_\text{BH-LMGO}&=+\Gamma_\text{D}+\Gamma_\text{E}+\Gamma_\text{G}-\Gamma_\text{H}-\Gamma_\text{K}-\Gamma_\text{O}-\Gamma_\text{BH+LMGO},\\
			\dot{N}_\text{LMGO-LMGO}&=+\Gamma_\text{C}-\Gamma_\text{G}-\Gamma_\text{I}-\Gamma_\text{LMGO+LMGO},\\
			\dot{N}_\text{BH-BH}&=+\Gamma_\text{F}+\Gamma_\text{H}-\Gamma_\text{Q}-\Gamma_\text{BH+BH},\\
			\dot{N}_\text{\BHprime-LMGO}&=+\Gamma_\text{I}+\Gamma_\text{J}+\Gamma_\text{K}+\Gamma_\text{R}-\Gamma_\text{S}-\Gamma_\text{U}-\Gamma_\text{\BHprime+LMGO},\\
			\dot{N}_\text{\BHprime-BH}&=+\Gamma_\text{O}+\Gamma_\text{P}+\Gamma_\text{Q}+\Gamma_\text{S}+\Gamma_\text{T}-\Gamma_\text{W}-\Gamma_\text{\BHprime+BH},\\
			\label{eq:BHPBHP}
			\dot{N}_\text{\BHprime-\BHprime}&=+\Gamma_\text{U}+\Gamma_\text{V}+\Gamma_\text{W}-\Gamma_\text{\BHprime+\BHprime}.
		\end{align}
		\label{differentialSystem}
	\end{subequations}
\end{widetext}

The overdots denote time derivative. This is a stiff problem and we employ backward differentiation 
formula methods to numerically solve it. We implement the FORTRAN package {\tt ODEPACK} \cite{ODEPACK}.
We note that for the binaries containing two compact objects i.e. the ones described by 
Eq.~\ref{eq:BHLMGO}-\ref{eq:BHPBHP}, we evolve their SMA axis to account for their 3rd-body hardening interactions.
The total number of such binaries at any time is given by the sum over all values of SMAs. For instance, for the 
BH-LMGO binaries $N_\textrm{BH-LMGO} = \Sigma_{i=1}^{100} N^{[i]}_\text{BH-LMGO}$ terms over the SMA bins.
Also, both the hard interaction rate of Eq.~\ref{interRate} and the exchange cross-section (see discussion 
in~\ref{sec:Exchange}), affecting the relevant exchange rate are evaluated including all 100 values of SMAs.
Eqs.~\ref{eq:LMGOSTAR}-\ref{eq:BHPSTAR} do not have SMA binning as 3rd-body interactions with other 
stars do not result in the binaries hardening. Nevertheless, tracking the evolving number of BH-star and LMGO-star 
pairs is essential as these are intermediate states towards the more interesting compact object binaries.

For the initial conditions of the differential system of Eqs.~\ref{eq:LMGO}-\ref{eq:BHPBHP} we take all BHs 
and LMGOs to be single objects. We consider each population to be uniformly 
distributed within their segregation volume. We discuss the evaluation of segregation volumes for different 
compact objects in Appendix~\ref{AppVelos}. The integration of the differential equations is proceeded along 
the segregation volume of the BHs, $V_{\textrm{BH}}$ because the BHs can only interact with the LMGOs 
in the intersection region of their segregation volumes. Subsequent interactions of these compact objects
 with star-star binaries will induce pairs including BHs and LMGOs through a cascade exchange processes 
 as in Fig.~\ref{diagram}. We also note that when compact objects are replaced by more massive ones through 
 exchanges (as e.g. in channels ``G'' of ``H'' for LMGOs), those objects are taken to remain inside the cluster. 
 We have estimated that their velocity after the exchange is still smaller than the typical escape velocity from
 the globular clusters. 
  
The number of star-star binaries is an initial condition in this work. These could be protobinaries\footnote{By 
the term ``protobinary'' here we mean a binary that was created when the cluster was form.} or could have been 
assembled via three body induced mechanism. To parametrize the number of 
star-star pairs in the core of a given globular cluster we take the number density of stars in the 
core  as $n_\text{star}={\rho_0\over m_*}$. The symbol $\rho_0$ is the central mass density of the cluster 
obtained directly from the globular cluster Harris catalog \cite{harris1996}. We take a fraction $f_\text{bin}$ of 
these stars to participate in star-star binaries and of them only a fraction $f_\text{hard}$ to be in 
hard binaries. Therefore, the number density of hard star-star pairs is given by $n_\text{star-star} = 
{1\over2}f_\text{hard}f_\text{bin}n_\text{star}$ with typical value for $f_\text{hard} \cdot f_\text{bin} = 0.05$. 
We take that to be our choice in this work. We probe the effect this fraction has on our results in 
Appendix~\ref{AppParameters}. The total number of these binaries inside the segregation volume of the 
BHs is equal to  $N_\text{star-star}=n_\text{star-star}V_\text{BH}$. Furthermore, we do not evolve the number 
of star-star binaries,  as their number is far greater than that of the compact objects and essentially constant.
	
We have included ``attenuation'' terms for the LMGO-LMGO, BH-LMGO \BHprime-LMGO, BH-BH, \BHprime-BH 
and \BHprime-\BHprime binaries that represent their mergers. These terms correspond to the coalescence of 
hard binaries through the process of hardening by 3rd-body interactions with stars \cite{Sesana:2015haa}. 
Given a type A-B binaries inside a cluster of number $N_{A-B}$, their merger rate is going to be,
\begin{equation}
\Gamma_\textrm{A+B}(t)= N_{\textrm{A-B}}^{a =
 a_{\textrm{GW}_{\textrm{A-B}}}}(t)/
T_{\textrm{GW}_{\textrm{A-B}}}, 
\end{equation} 
which evolves with time as the number of binaries of type A-B with a SMA of $a = 
a_{\textrm{GW}_{\textrm{A-B}}}$, $N_{\textrm{A-B}}^{a =
 a_{\textrm{GW}_{\textrm{A-B}}}}$ evolves with time as well. 

\subsection{Binary-single exchange cross section and exchange rates}
\label{sec:Exchange}
The exchange cross section for an object of mass $m_3$ to substitute an object of mass $m_1$ in a hard binary 
system $1-2$ is given by, 
\begin{equation}
\Sigma_\text{ex}={G(m_1+m_2+m_3)a\over v_{3,1-2}^2}\times f(m_1,m_2,m_3).  
\end{equation} 
$a$ here is the SMA of $1-2$ before the exchange and $v_{3,1-2}$ the relative velocity of object 3 to the binary 
calculated at infinity. As the $1-2$ is a hard binariy the exchange cross section contains only the gravitational focusing term.
This is averaged over orbital eccentricities following a thermal distribution $P(e)de=2ede$. 
The form factor, $f(m_1,m_2,m_3)$ is equal to \cite{Heggie:1996bs},
\begin{widetext}
\begin{eqnarray}
f(m_1,m_2,m_3)&=&{m_3^{7/2}(m_2+m_3)^{1/6}\over (m_1+m_2)^{1/3}(m_1+m_3)^{5/2}(m_1+m_2+m_3)^{5/6}} \\
&\times&\exp(3.70+ 7.49x-1.89y-15.49x^2-2.93xy-2.92y^2 +3.07x^3+ 13.15x^2y-5.23xy^2+3.12y^3), \nonumber 
\end{eqnarray}
\label{eq:massDepFactor}
\end{widetext}
with $x=m_1/(m_1+m_2)$, $y=m_3/(m_1+m_2+m_3)$ and valid within the point-mass approximation. 

The corresponding velocity averaged exchange rate density for $1-2\to3-2$, assuming a Maxwell-Boltzmann distribution 
for the relative velocity, is given by,
\begin{equation}
\gamma_\text{ex}=\sqrt{6\over\pi}n_{1-2}n_3{G(m_1+m_2+m_3)a\over\sigma_{3,1-2}}\times f(m_1,m_2,m_3).
\label{gammaEX}
\end{equation}
The number densities $n_{1-2}$ and $n_3$ depend on the radial position inside the cluster.
The exchange rates that enter Eqs.~\ref{eq:LMGO}-\ref{eq:BHPBHP}, are just,
\begin{equation}
\Gamma_{\textrm{ex}} = \int \gamma_{\textrm{ex}} dV = \gamma_\text{ex}\times\min(V_{1-2},V_{3}).
\label{eq:GammaTotEx}
\end{equation}
The integration is done over the portion of volume of the cluster where $1-2\to3-2$ exchanges occur. 
We consider the $1-2$ and $3$ to be distributed uniformly within their respective segregation volumes 
$V_{1-2}$ and $V_3$. 

\begin{figure}
	\centering
	\includegraphics[width=0.5\textwidth]{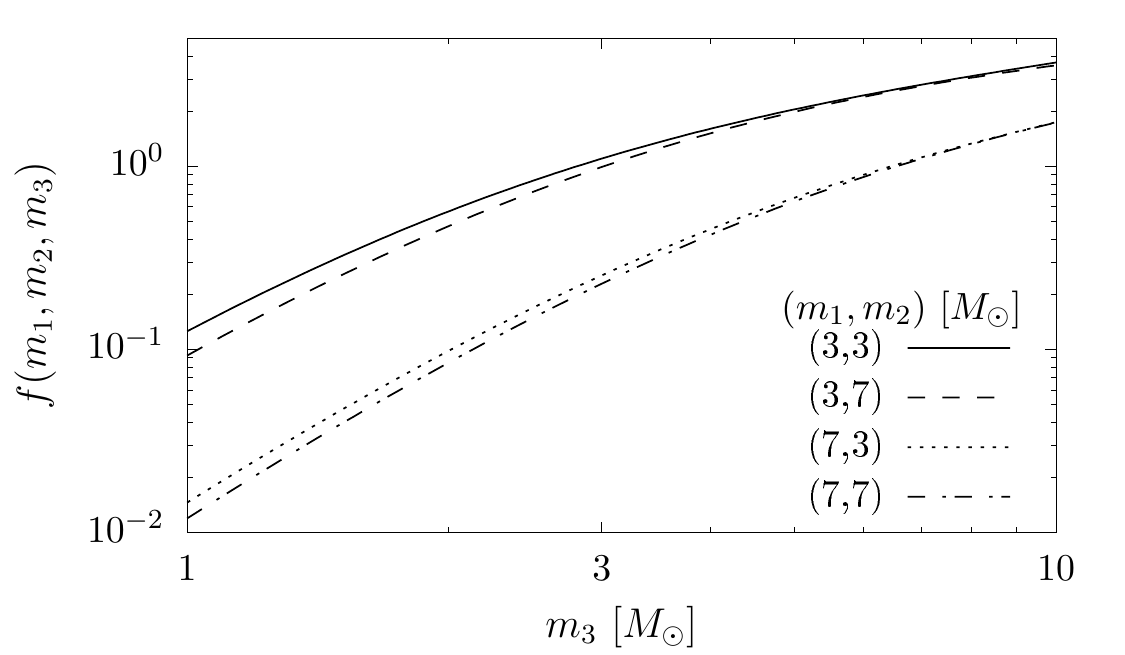}
	\caption{The form factor $f(m_1,m_2,m_3)$ as a function of the third mass $m_{3}$ 
	and for four choices of binary masses $(m_{1}, m_{2})$. It is assumed that the $m_1$ is exchanged by $m_3$.}
	\label{massFactor}
\end{figure}

In Fig.~\ref{massFactor} we show $f(m_1,m_2,m_3)$ for certain choices of masses. We note 
that the exchange process $1-2\to3-2$ has a high probability of occurring as the third object 
approaching the binary has a mass larger than both of the binary components. On the other hand, 
if $m_3<\min(m_1,m_2)$ the cross section drops and the probability of exchange becomes essentially 
zero when the $m_3<m_1/\sqrt{10}$. Furthermore, there is a strong dependence between the mass of the intruder 
and of the exchanged object, while a weak dependence on the mass of the remained object. Therefore, 
since the mass of stars is typically $m_* < 1M_\odot$, 
and the masses of BHs and LMGOs is greater than 3$M_\odot$, we neglect 
exchanges in which a star substitutes an LMGO or a BH.

\subsection{The impact of second generation black holes}

We consider the situation in which a pair of first generation black holes 
BH of mass 10 $M_{\odot}$ each coalesces to give a second generation black hole \BHprime \,
with a mass of 19 $M_{\odot}$. In realistic situations where the masses of the merging
BHs are not equal, the product \BHprime obtains a GW kick and may escape from the cluster. 
In Appendix~\ref{App:2ndGenBHsKicks} we describe in full detail how we treat this kick and 
how we calculate the fraction of the retained \BHprime \, in each globular cluster.
 
\begin{figure}
\centering
\includegraphics[width=0.5\textwidth]{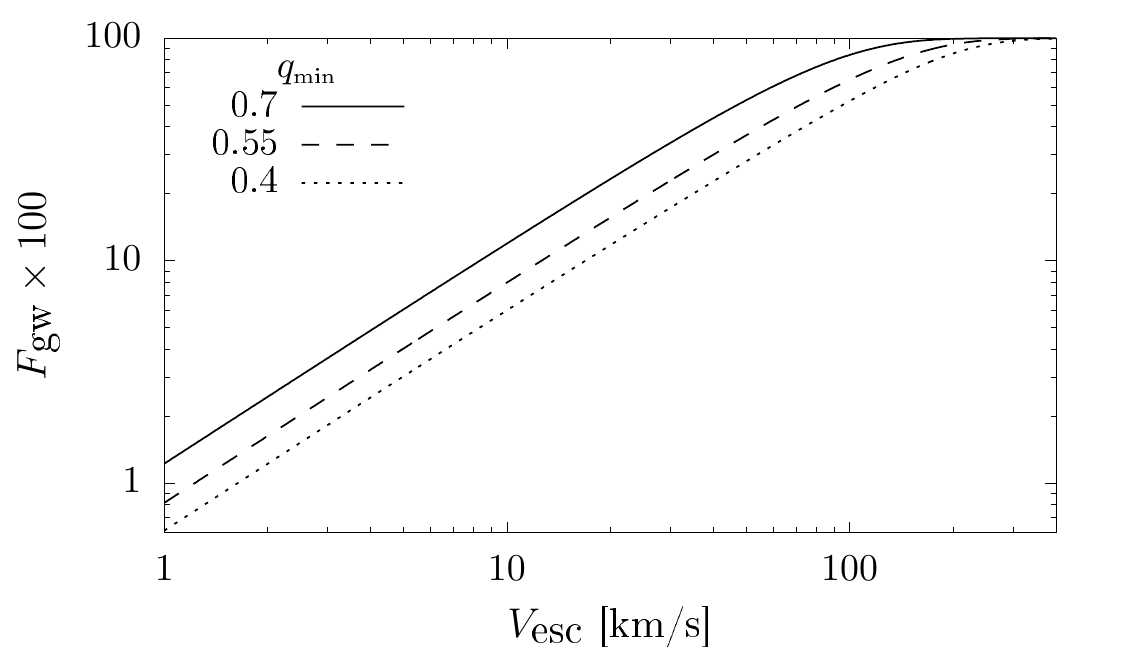}
\caption{The retention fraction of second generation \BHprime merger remnants as a function 
of the escape velocity from their host cluster and under the assumption of a uniform mass ratio 
distribution. The lines are for three values of the smaller mass ratio $q_\text{min}$. The labels 
denote the value of $q_\text{min}$.}
\label{FgwPLOT}
\end{figure}

We find that the retention fraction of second generation \BHprime \, in a cluster with escape 
velocity of 50km/s (a typical value for large GCs), is approximately $20\%$. For an 
escape velocity above 400km/s the fraction saturates to $100\%$. In Fig.~\ref{FgwPLOT} 
we show the retention fraction $F_{\textrm{gw}}$ for \BHprime \, objects versus the escape velocity 
from a cluster. The $q_\text{min}$  refers to the minimum value of mass ratio between 
the merging first generation black holes for a realistic BH mass spectrum. The $q_\text{min}=0.55$ 
curve that is also our reference choice gives a rough estimate of the retention fraction of 
second generation black holes, which we find that can be parametrized well as,
\begin{equation}
F_\text{gw}(V_\text{esc})\simeq\left({V_\text{esc}\over1.2\text{km/s}}\right)^{0.97}\%,
\ \text{for}\ V_\text{esc}<100\text{km/s}.
\label{retFracGW}
\end{equation}
This relation holds true to within a few percent for the vast majority of Milky Way type GCs. 

The total retention fraction in Eq.~\ref{retFracGW} controls the amount of second 
generation black holes \BHprime, which remain in the cluster and are imported in our 
Fig.~\ref{diagram}. The corresponding creation rate of \BHprime \; is $F_\text{gw}\times 
\Gamma_\text{BH+BH}$, where $\Gamma_\text{BH+BH}$ is the merger rate of BH-BH pairs.
	
\section{Results}
\label{secResults}
	
The solution of the differential system in Eqs.~\ref{eq:LMGO}-\ref{eq:BHPBHP}, yields 
the time evolution of the population sizes inside $V_\text{BH}$. In this section we 
present the results of this model.
	
\begin{figure*}[t]
\centering
\includegraphics[width=0.49\textwidth]{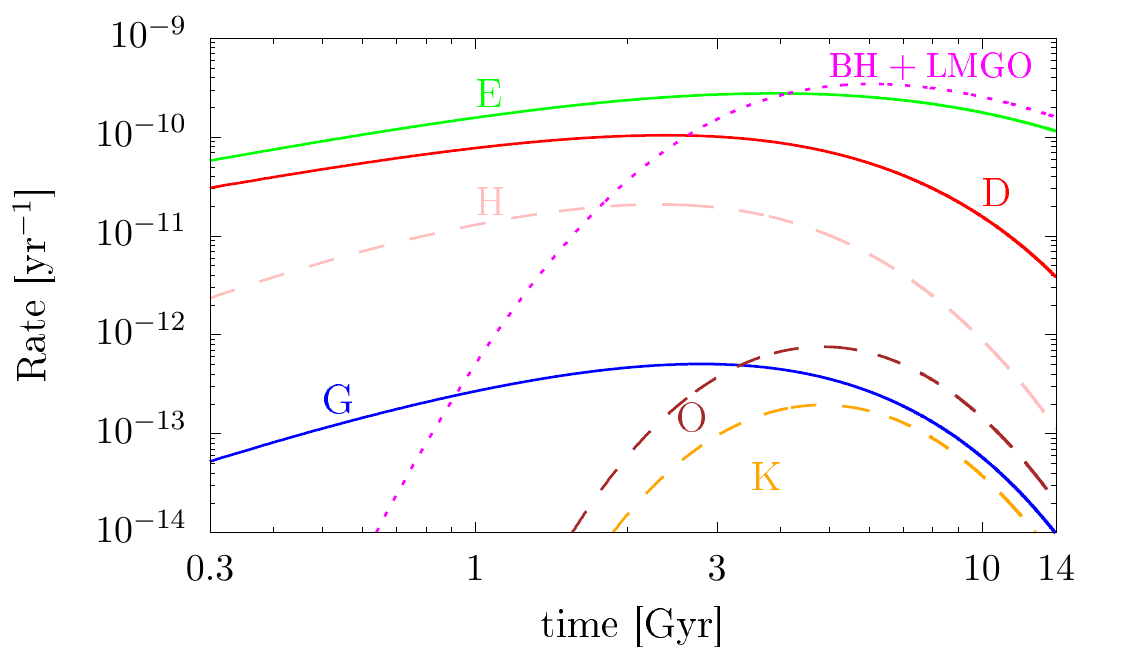}
\includegraphics[width=0.49\textwidth]{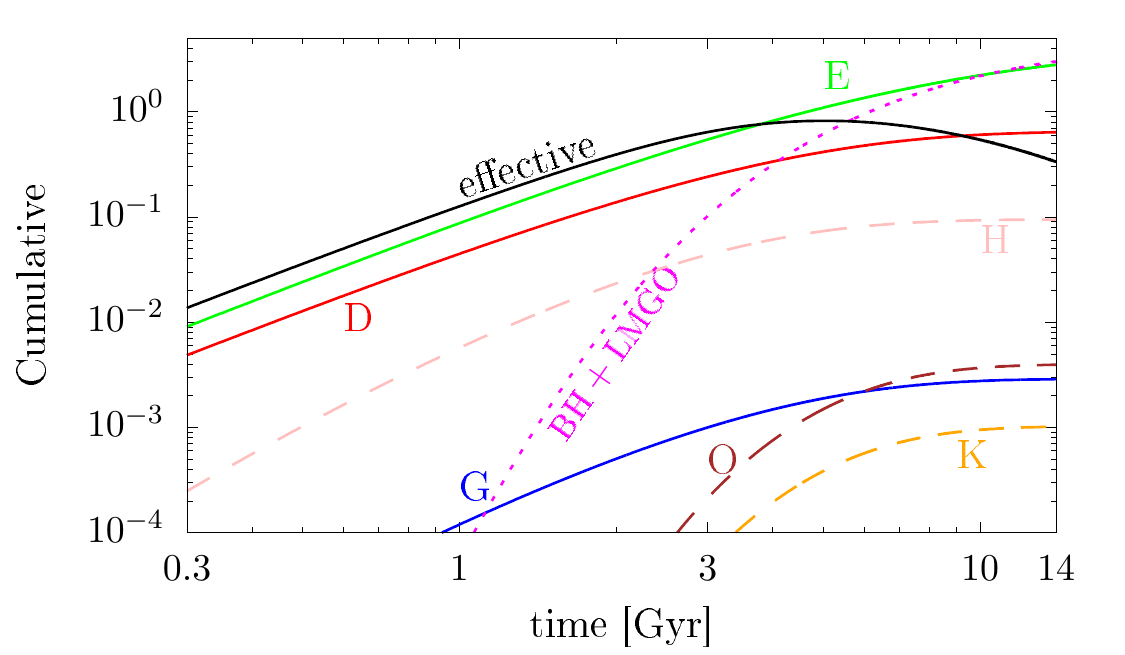}
\caption{{\it Left panel:} The exchange and merger rates associated with the BH-LMGO binaries (see Fig. 
\ref{diagram}). Solid lines represent exchange channels that increase the number of BH-LMGO binaries inside the 
cluster and dashed lines  channels reducing it via exchanges.  
{\it Right panel:} The corresponding cumulative number of BH-LMGO pairs created or annihilated in the 
history of the cluster as well as the effective number of BH-LMGO pairs as a function of time. The black 
line denoted as ``effective'' shows the evolution of the number of BH-LMGO binaries. To produce these 
curves we used the properties of Terzan 5 globular cluster.}
\label{fig:1genBH_LMGO}	
\end{figure*}	
	
\subsection{The depletion of BHs and LMGOs}
	
Single BHs and LMGOs first create binaries with stars which then they exchange with more massive objects. 
While single BHs are more numerous than the LMGOs, due to their larger mass their numbers deplete faster. 
LMGOs create stable binaries with other LMGOs or BHs when the number of single BHs is significantly reduced. 
Processes like ``H'' in Fig.~\ref{diagram} in which a BH-LMGO pair becomes a BH-BH pair and an LMGO is freed 
decelerates somewhat the rate by which the population of single LMGO depletes. This also explains why BH-BH 
mergers grow in numbers earlier than mergers involving an LMGO. BHs tend to coalesce with LMGOs in the late 
stages of a cluster after the binary BH population has settled down. 

\subsection{The formation of BH-LMGO binaries inside globular clusters}
\label{sec:BHLMGObin}

Binaries of first generation black holes BH with low mass gap objects are a class of binary systems that 
form dynamically inside clusters. These binaries are seeded from three binary populations, LMGO-star 
pairs from channel ``D'',  BH-star pairs from channel ``E'' and LMGO-LMGO pairs via channel ``G'' 
(see Fig.~\ref{diagram}). However, they can also be depleted by exchange interactions with first 
generation black holes via channel ``H'' or when second generation black holes  \BHprime are formed and cause 
exchange interactions ``K'' and ``O''. In Fig.~\ref{fig:1genBH_LMGO}, we show the evolution with time of 
those rates (left panel) as the number of each of the individual objects and of the binaries involved changes 
with time. We use the environmental parameters of Terzan 5 globular cluster to produce these lines. 
Solid lines represent positive rate terms increasing the number of BH-LMGO binaries and dashed lines 
negative terms reducing it.  

The two main pathways in the formation of BH-LMGO pairs are the formation first of BH-star pairs, followed by the 
exchange of the star with a single LMGO (B $\rightarrow$ E $\rightarrow$ BH-LMGO) and the formation of a 
LMGO-star pair first, followed by an exchange interaction with a BH (A $\rightarrow$ D $\rightarrow$ BH-LMGO) 
\footnote{We assume that LMGO-star interactions with BHs always lead to the creation of a BH-LMGO pair and 
not to BH-star ones.}. Channel G is very rare as there aren't many LMGO-LMGO pairs in the first place. 
Even though, there are many ways to form a BH-LMGO binary, their numbers inside clusters are small. 
It is difficult to form single LMGOs. Also, many of these objects are at orbits inside the cluster but beyond 
the core dominated by BHs and do not interact with them. These prove more important factors in the suppression of the rates of 
BH-LMGO formed, compared to the occasional exchange interactions the BH-LMGO binaries can have with 
other black holes leading to LMGOs getting replaced by first or second generation black holes (dashed lines 
in Fig.~\ref{fig:1genBH_LMGO}). In the right panel of Fig.~\ref{fig:1genBH_LMGO}, we show the cumulative 
number of BH-LMGO binaries formed over time from each specific channel (in solid lines) and also the cumulative 
number of BH-LMGO binaries destroyed by exchanges with black holes from each specific channel (in dashed lines). 

In addition, in Fig.~\ref{fig:1genBH_LMGO}, we present the evolution with time of the total BH-LMGO merger rate 
and number of BH-LMGO mergers inside Terzan 5 (dotted lines at the left and right panels respectively). We allow 
those numbers to be non-integers as we will later present calculations involving other Milky Way cluster environments 
and will sum the contribution from many clusters to evaluate their typical averaged contribution to the total rates of 
BH-LMGO mergers in the Universe. 
\begin{figure*}[ht]
\centering
\includegraphics[width=0.49\textwidth]{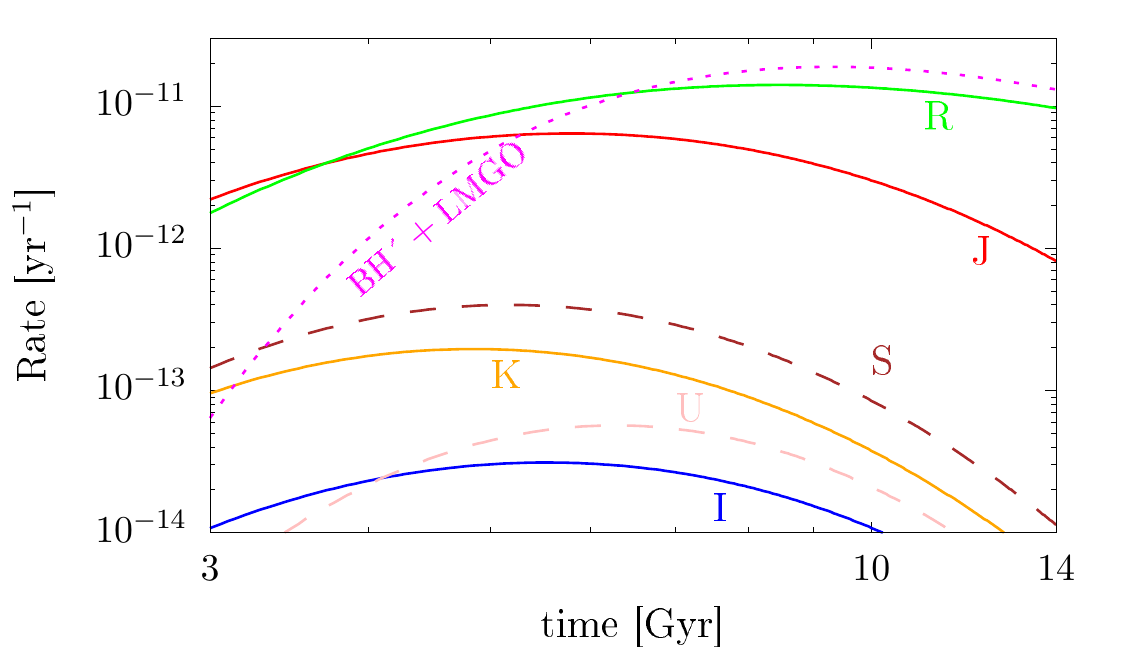}
\includegraphics[width=0.49\textwidth]{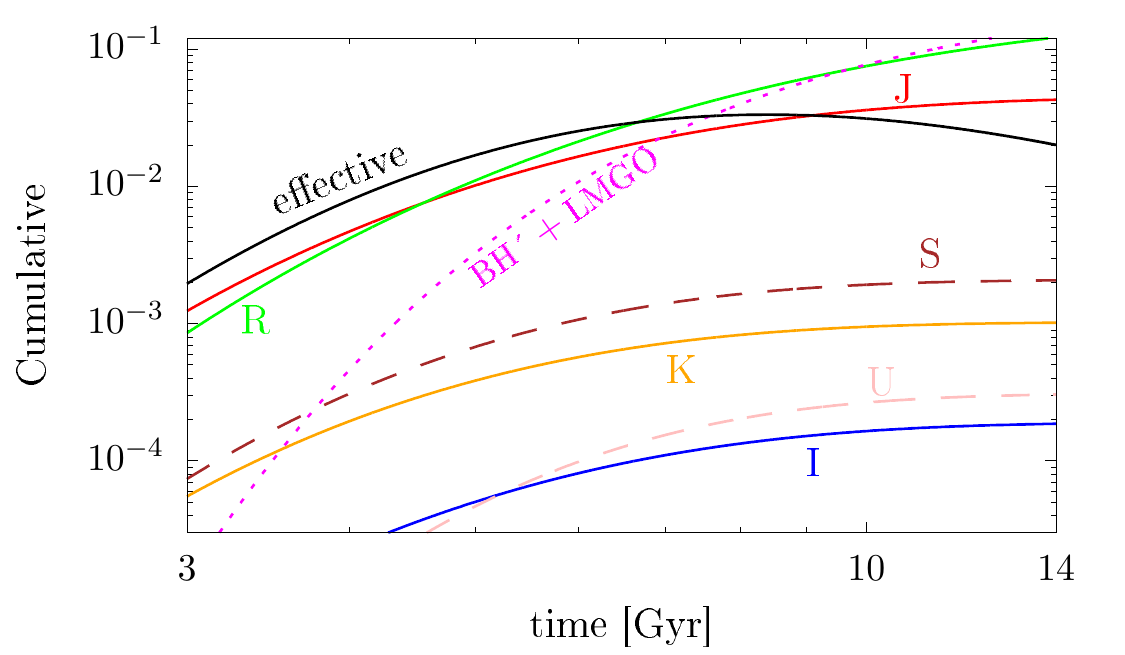}
\caption{{\it Left panel:} The exchange and merger rates associated with the \BHprime-LMGO binaries (see 
Fig.~\ref{diagram}). {\it Right panel:} The corresponding cumulative number of \BHprime-LMGO pairs created 
or disrupted in the history of the Terzan 5 cluster as well as the effective number of \BHprime-LMGO pairs as 
a function of time.}
\label{fig:2genBH_LMGO}	
\end{figure*}
Toward the end of the cluster's history presented, the number of BH-LMGO pairs (line denoted as 
``effective'') drops as mergers occur. 
	
Since second generation black holes \BHprime coming from the BH-BH mergers are formed inside the clusters, we also 
include their impact both in disrupting BH-LMGO binaries but also in creating the more rare \BHprime-LMGO pairs, 
i.e. binaries where the black hole is a second generation one. Those are presented in Fig.~\ref{fig:2genBH_LMGO}. 
We follow the same notation as for Fig.~\ref{fig:1genBH_LMGO}, where positive rates for the formation of such binaries 
come from channels ``I'', ``J'', ``K'' and ``R'', and negative disruption rates due to exchanges of LMGOs by 
more massive black holes from channels ``S'' and ``U''. The number of \BHprime-LMGO binaries is dominated by the 
number of LMGO-stars having exchange interactions with \BHprime objects (A $\rightarrow$ J $\rightarrow$ \BHprime-LMGO)
and by \BHprime-star binaries having exchange interactions with single LMGOs (M $\rightarrow$ R $\rightarrow$ \BHprime-LMGO).
However, we find that such binaries are far less common to exist and merge by about a factor of $\sim$30, as second 
generation black holes are far rarer. If their number inside dense stellar environments is enhanced so will their respective binaries. 

\begin{figure*}
\centering
\includegraphics[width=\textwidth]{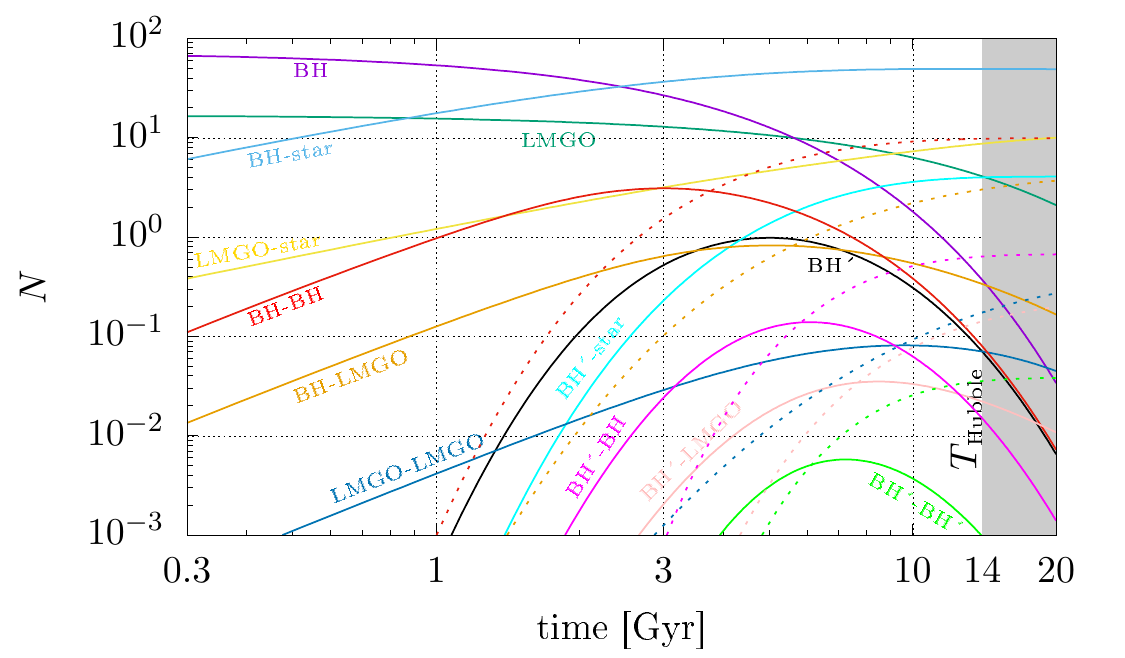}
\caption{The time evolution of all single compact objects (LMGO, BH, \BHprime) and binary types that we consider 
in our model (Fig.~\ref{diagram}) as well as their corresponding cumulative number of mergers in dotted 
lines. These numerical results have been obtained in the core of our reference cluster Terzan 5. 
The gray shaded region represents the region to the right of Hubble time which is taken to be $14$Gyr.}
\label{fig:prototypical_All}
\end{figure*}

In Fig.~\ref{fig:prototypical_All}, we show the evolution of the numbers of single compact objects LMGO, BH and \BHprime
as well as the evolution of number of those compact objects with each other and with stars inside Terzan 5 globular 
cluster. For each binary type and for each type of interaction effecting those numbers we have taken into account the 
relevant segregation volumes of the objects involved as described in  Eqs.~\ref{eq:3rdBodyGamma} and~\ref{eq:GammaTotEx}. 
As can be seen
the formation of BH-LMGO binaries is somewhat delayed by comparison to BH-BH as LMGOs can be replaced by BHs
via exchanges.  Most single BHs are in BH-star and BH-BH pairs. Only after most BHs are in these binaries the number
of BH-LMGO binaries peaks. LMGO-LMGO binaries are even more rare and require for the number of single BHs to 
be very suppressed before they can remain in stable binaries that can then harden via 3rd-body interactions with stars. 

\BHprime-LMGO systems are delayed due to the required merger of BH+BH $\rightarrow$ \BHprime. Yet, once 
those second generation black holes are produced, the relevant cross-section for them to interact with other binaries 
is so large that binaries with a \BHprime as a member form very fast as seen by the rapidly increasing cyan 
(\BHprime-star), rose (\BHprime-BH), pastel (\BHprime-LMGO) and green (\BHprime-\BHprime) lines. 
	
\subsection{The formation of black hole binaries with a second generation black hole}

\begin{figure*}
\centering
\includegraphics[width=0.49\textwidth]{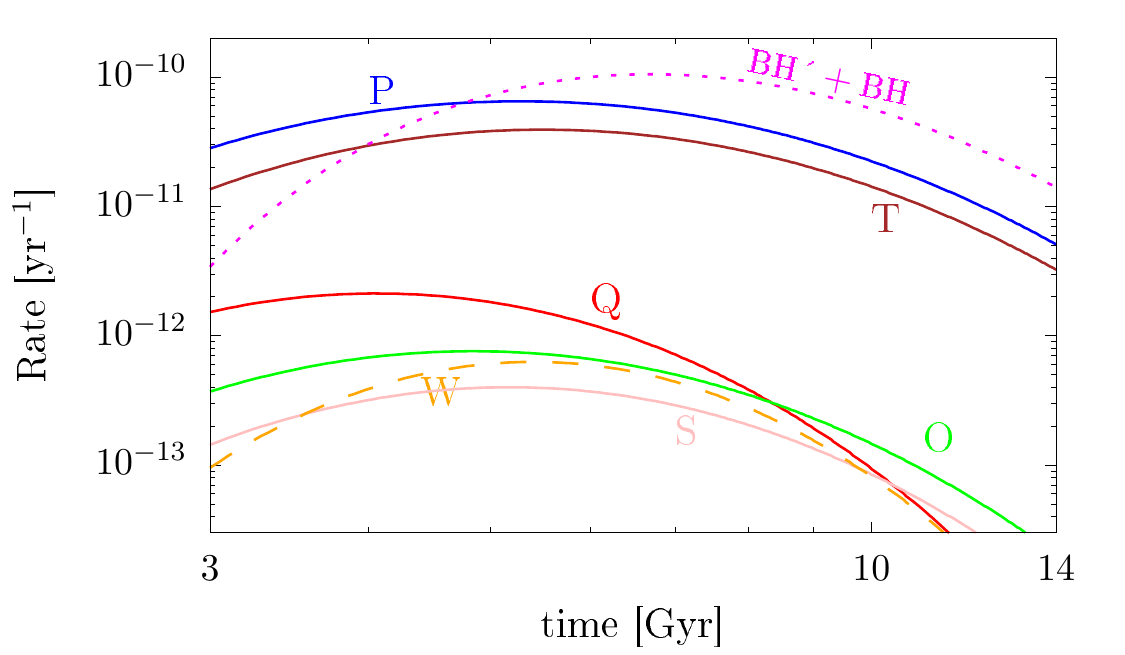}
\includegraphics[width=0.49\textwidth]{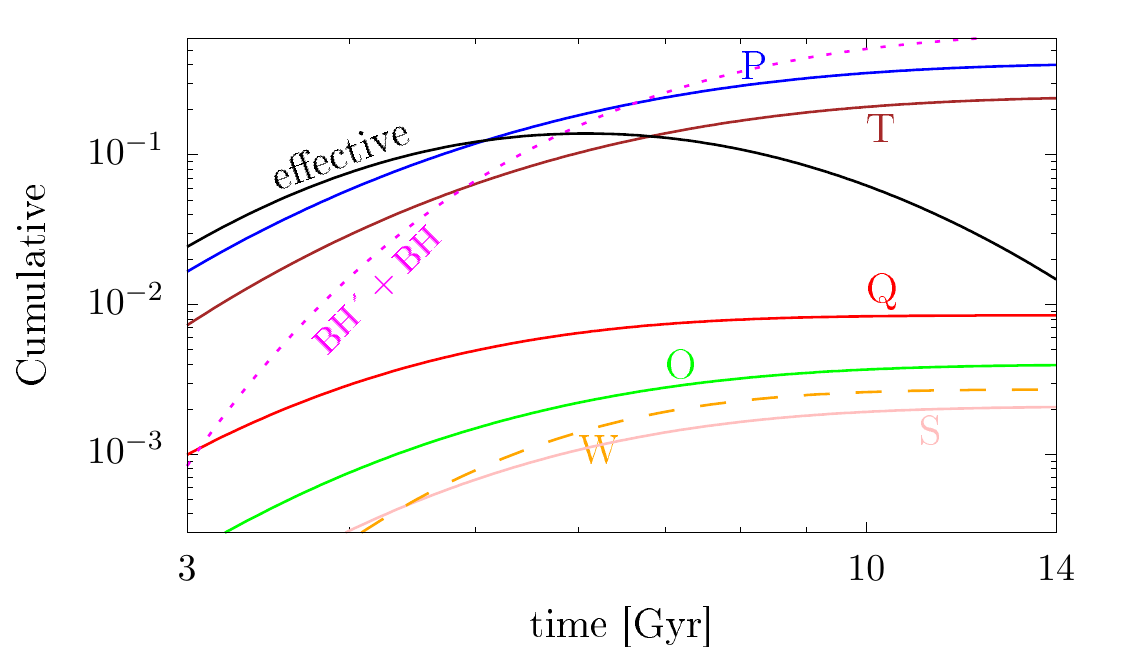}
\caption{{\it Left panel:} The exchange and merger rates associated with the \BHprime-BH binaries (see Fig.~\ref{diagram}). 
{\it Right panel:} The corresponding cumulative number of \BHprime-BH pairs created or annihilated in the history of 
Terzan 5 as well as the effective number of \BHprime-BH pairs as a function of time.}
\label{fig:2nd1stBBH}
\end{figure*}

In Fig.~\ref{fig:prototypical_All}, we also present the binaries formed containing a first generation and a second
generation black hole. Those are created via the channels ``O'', ``P'', ``Q'', ``S'' and ``T'' of Fig.~\ref{diagram} 
and are reduced just when a second \BHprime replaces the first generation black hole (channel ``W'').  In 
Fig.~\ref{fig:2nd1stBBH}, we show the evolution of those rates and of the relevant cumulative numbers.  The most 
important channels are those involving the binaries containing a star and a black hole which is then replaced by a  
second black hole, (B $\rightarrow$ P $\rightarrow$ \BHprime-BH and M $\rightarrow$ T $\rightarrow$ \BHprime-BH).	

While rare, these binaries still lead to \BHprime-BH mergers that are only about a factor of $\simeq$30 less common than 
the regular BH-BH merger events, i.e. two first generation black holes merging and thus can provide a detectable rate
of events. 
	
\subsection{Build-up of the LMGO population}

\begin{figure*}
\centering
\includegraphics[width=\textwidth]{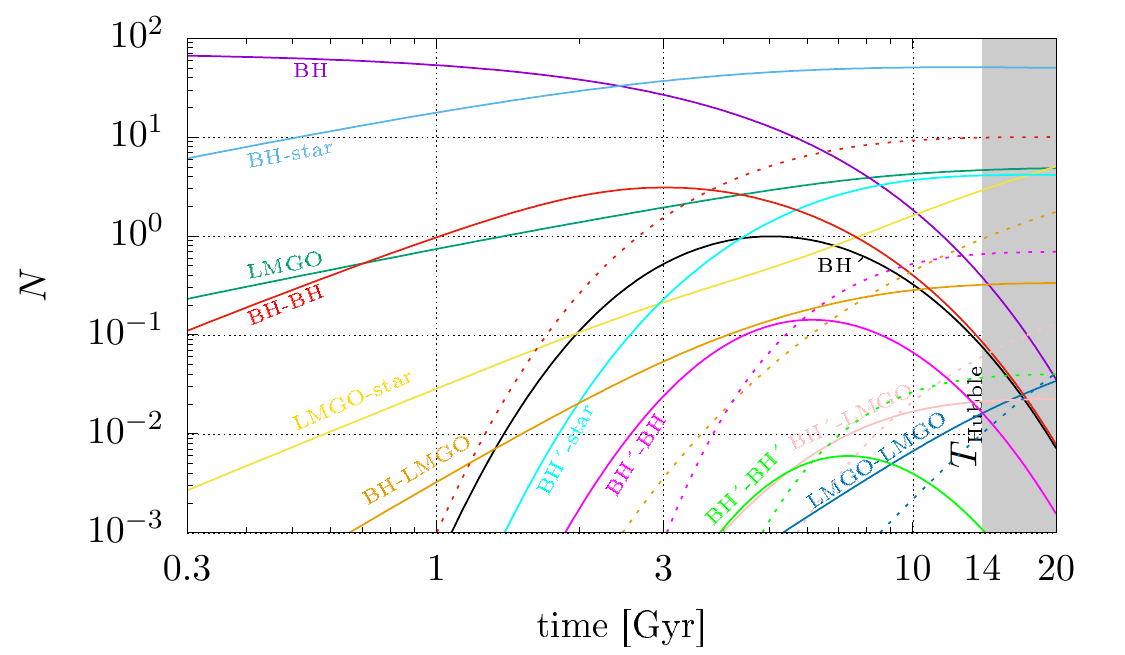}
\caption{As in Figure~\ref{fig:prototypical_All} but for a population of LMGOs that gradually builds up (solid green line). 
The time evolution of all single compact objects and binary types considered in this work is shown in solid lines. In dotted 
lines we show the corresponding cumulative number of mergers.}
\label{fig:noLMGO_plot}
\end{figure*}

A major assumption that we made in the above mentioned results is that there is an initial number of LMGOs 
inside the cluster that can dynamically interact with other objects already in the first time-step. 
Unlike first generation BHs that form in the very first $\sim$10 million years of the clusters's existence, 
LMGOs depending on their origin may require much longer time to form.
In this section instead, we make the assumption that the population of $3M_\odot$ LMGOs is created via NS-NS 
mergers inside the GCs and there is a gradual build-up of the LMGO's population. We take the initial population 
of LMGOs to be zero, $N_\text{LMGO}(t=0)=0$ and consider 
a source term into the equation for the evolution of LMGOs represented by the NS+NS merger rate.  Specifically, 
we rewrite Equation~\ref{eq:LMGO} adding the relevant source term,
\begin{eqnarray}
\dot{N}_\text{LMGO}&=&F_{\textrm{gw}}\Gamma_{\textrm{NS+NS}}-\Gamma_{\textrm{A}}-\Gamma_{\textrm{C}} 
- \Gamma_{\textrm{E}}+\Gamma_{\textrm{G}}+\Gamma_{\textrm{H}}+\Gamma_{\textrm{I}} \nonumber \\
&+&\Gamma_{\textrm{L}} +\Gamma_{\textrm{O}} - \Gamma_{\textrm{R}} + \Gamma_{\textrm{S}} + \Gamma_{\textrm{U}},
\end{eqnarray}
$F_{\textrm{gw}}$ is taken to be 1 and thus the source term is,
\begin{equation}	
\Gamma_{\textrm{NS+NS}}(t)=N(t)\Theta(N(t))\Theta(T_\text{Hub}-t){f_e\left(T_\text{Hub}-t\right)\over T_\text{Hub}-t}
\end{equation}
where $T_\text{Hub}$ is the Hubble time, $\Theta$ is the Heaviside function and $N(t)=\left(N_\text{NS}/2-N_\text{LMGO}(t)\right)$ 
is the number of NS-NS hard pairs at time $t$. 

We point out that for this work the time $t=0$ is the moment that both the first generation of BHs and the 1 
$M_{\odot}$ stars have been formed and exist in binaries inside the cluster. We also assume that the segregation 
of BHs in the cluster's core has taken place. 1 $M_{\odot}$ stars still require $\simeq 40$ Myrs to form from 
contraction of their initial gas clouds and thus follow the formation of first generation BHs which only require $O(10)$ 
Myrs altogether. The segregation of BHs inside the core also requires $O(10)$ Myrs. All these are significantly smaller 
timescales compared to the exchange ones at $t \simeq 0$. Thus exchanges happen after those phases in the 
history of a cluster have taken place.

In Fig.~\ref{fig:noLMGO_plot}, we show the cumulative number of compact objects, their binaries and of merger 
events for Terzan 5 as they all evolve. This 
can be compared to Figure~\ref{fig:prototypical_All} that did not include the gradual build-up of LMGOs. 
With the obvious exception of changing the number of binaries containing LMGOs, the effect of a gradual built up of LMGOs 
is entirely insignificant for all the other types of binaries. All channels in Fig.~\ref{diagram} and 
Equations~\ref{eq:LMGO}-\ref{eq:BHPBHP} describing LMGOs have a very small effect on the evolution of 
non-LMGO binaries.

We point out that, even after including this major change on the initial conditions of the LMGOs number, the number of 
BH-LMGO and the \BHprime-LMGOs mergers in the history of the cluster is only suppressed by a factor of 
four and two respectively compared to the case were we included LMGOs from the first time-step. 
The time-demanding build-up included in this section represents a slow pathway in the creation of LMGOs 
and our results in section~\ref{sec:BHLMGObin} an ``instant'' one. Thus in combination our results provide a wide
width of possibilities. Our results of modeling the dynamical interactions of LMGOs with other objects inside 
dense stellar clusters are quite generic and independent to the exact origin of the LMGOs. As long as their 
number in a stellar environment does not exceed that of first generation BHs our results can be applicable to 
wide range of possible explanations on the BH-LMGO events as GW190814.

\subsection{Average Cumulative Cases and merger rates}

In Figs.~\ref{fig:1genBH_LMGO} to~\ref{fig:noLMGO_plot} we used for reference Terzan 5 cluster as it is a 
massive one that can contain significant numbers of BH, \BHprime and LMGOs. While Terzan 5 is an instructive 
example, we care for the averaged rate of mergers in globular clusters. For that we use the 106 Milky Way clusters 
for which we have reliable data of their mass profile. Those rejected are the smallest mass ones,

We evaluate the average cumulative number of mergers per cluster for the six binary species we consider, 
i.e. BH-LMGO, LMGO-LMGO, BH-BH, \BHprime-LMGO, \BHprime-BH and \BHprime-\BHprime. While in 
the Figs.~\ref{fig:1genBH_LMGO} to~\ref{fig:noLMGO_plot} we plotted results all the way to the Hubble time, 
the typical age of cluster is a few Gyrs smaller. For that reason we calculate the cumulative number of mergers at 
10 Gyr of evolution time. In Table~\ref{tab:values2}, we show the expected number of mergers in the history of a
 variety of Milky Way clusters for each type of binary. Also, (at the last line) we give the average number of mergers 
 per cluster.  We assume no delay in the initial population of LMGOs.

\begin{table*}
	\centering
	\begin{tabular}{c| c c c c c c c c}
		\hline
		GC & $N^\text{init.}_\text{BH}$ & $N^\text{init.}_\text{LMGO}(V_\text{BH})$ & BH+LMGO & LMGO+LMGO & BH+BH & \BHprime+LMGO & \BHprime+BH & \BHprime+\BHprime \\
		& & & & $\times10^{-2}$ & & $\times10^{-2}$ & $\times10^{-1}$ & $\times10^{-2}$ \\
		\hline
		\hline
Terzan 5 & 73 & 16 & 2.19 & 9.24 & 9.15 & 6.36 & 4.22 & 1.75 \\
NGC 104 & 499 & 109 & 4.03 & 6.90 & 44.0 & 2.30 & 7.88 & 2.08 \\
NGC 1851 & 82 & 18 & 2.31 & 8.86 & 10.6 & 5.98 & 4.77 & 2.02 \\
NGC 6266 & 415 & 105 & 6.18 & 15.1 & 51.4 & 16.1 & 24.9 & 12.6 \\
NGC 6293 & 11 & 2 & 0.44 & 3.24 & 1.19 & 0.845 & 0.258 & 0.07 \\
NGC 6440 & 119 & 29 & 3.71 & 18.1 & 16.4 & 12.7 & 9.85 & 4.86 \\
NGC 6441 & 788 & 208 & 12.8 & 45.5 & 106 & 45.7 & 65.8 & 45.2 \\
NGC 6522 & 9 & 1 & 0.395 & 2.91 & 1.08 & 0.767 & 0.233 & 0.07 \\
NGC 6624 & 8 & 1 & 0.251 & 1.87 & 0.654 & 0.368 & 0.112 & 0.02 \\
NGC 6626 & 99 & 21 & 1.67 & 5.49 & 12.2 & 2.86 & 3.66 & 1.31 \\
NGC 6681 & 9 & 1 & 4.62 & 3.42 & 1.30 & 0.115 & 0.347 & 0.14 \\
NGC 7078 & 540 & 122 & 64.7 & 13.2 & 64.4 & 1.24 & 25.2 & 11.3 \\
		\hline
average & 82 & 21 & 0.57 & 1.9 & 4.3 & 1.8 & 2.0 & 1.3 \\
		\hline
	\end{tabular}
	\caption{The cumulative number of mergers for each of the six binary species 
	at 10 Gyr of evolution. In the first 12 rows present the clusters with the highest 
	contribution. The last row shows the average cumulative number of mergers averaged 
	over the ensemble of 106 Milky Way globular clusters. 
	}
	\label{tab:values2}
\end{table*}

BH-BH mergers provide the most abundant class of mergers in globular clusters. Such binaries are build from 
BH-LMGO and BH-star binary exchange interactions with single BHs. They are also well measured by LIGO-Virgo and 
can provide a reference for the more exotic channels. In terms of rates we get that there are $5.7\times 10^{-11}$ yr$^{-1}$ 
per cluster BH-LMGO mergers and $1.8\times 10^{-12}$ yr$^{-1}$ per cluster \BHprime-LMGO mergers.
By comparison, for BH-BH mergers which constitute the majority of 
LIGO's binary black hole merger events, we get $4.3\times 10^{-10}$ yr$^{-1}$ per cluster mergers.  
We do find though that globular clusters can give $2.0\times 10^{-11}$ yr$^{-1}$ 
 \BHprime-BH and $1.3\times 10^{-12}$ yr$^{-1}$ per cluster \BHprime-\BHprime mergers. 
Such rates allow for LIGO-Virgo to see both BH-LMGO and \BHprime-BH  in a small sample of its events, 
as it may already have with GW190814 for the BH-LMGO case. We also find a rate of $1.9\times 10^{-12}$ yr$^{-1}$ 
per cluster for LMGO-LMGO binaries which we expect will be observable events as the LIGO-Virgo sensitivity develops. 

To get a conservative estimate for local BH-LMGO binary merger rate, we take the local number density 
of globular clusters to be constant at $n_{GC}(z) = 0.77\times10^{9}$ Gpc$^{-3}$ \cite{Rodriguez:2016kxx}. We 
also allow for an evolving with redshift globular cluster density, as $n_{GC}(z) = 0.77 \times 10^{9}  \cdot E(z)^{3}$ 
Gpc$^{-3}$ \footnote{$E(z)=\sqrt{\Omega_M\cdot(1+z)^3+\Omega_{\Lambda}}$, with 
$\Omega_{M}=0.3$ and $\Omega_{\Lambda}=0.7$, based on Planck data (2015) \cite{Ade:2015xua}.}. Assuming the Milky 
Way sample to be representative of the averaged local Universe up to redshift $z$ of 1, we estimate the cumulative merger 
rate $\mathcal{R}_c(z=1)$  \cite{Ye:2019xvf, Rodriguez:2016kxx}, 
 \begin{equation}
    \mathcal{R}_c(z=1)=\int_{0}^{z=1} dz'\ \langle\Gamma_{GC}(z')\rangle\ n_{GC}\ {dV_c\over dz'}\ (1+z')^{-1}.
\end{equation}
where $V_c$ is the comoving volume. We get a merger rate density of $\Gamma_\text{BH-LMGO} = \mathcal{R}_c(z = 1)/V(z=1)  
\simeq (2-8)\times10^{-2} \text{Gpc}^{-3}\text{yr}^{-1}$. 
If instead, we use the hypothesis that the number of LMGOs was build-up via NS-NS mergers then that rate density 
becomes  $\Gamma_\text{BH-LMGO}\simeq (0.4-1.3)\times10^{-2} \text{Gpc}^{-3}\text{yr}^{-1}$.
By comparison, the experimentally inferred merger rate density of GW190814-like events is $(1-23)\text{Gpc}^{-3}\text{yr}^{-1}$
Our predicted rates tend to be lower by an order of magnitude compared to the LIGO-Virgo values. As we 
pointed out if more LMGOs are produced via a non-NS channel, those rates will be enhanced. Also, other environments 
as nuclear clusters in galaxies whose population properties are less well-understood, could enhance the BH-LMGO merger 
rate density. If the retention fractions of BHs and NSs are increased by similar factors, then all the merger rates 
would grow, while the ratio of between the different types of merger events would stay approximately the same. 

\begin{figure}
	\includegraphics[width=0.5\textwidth]{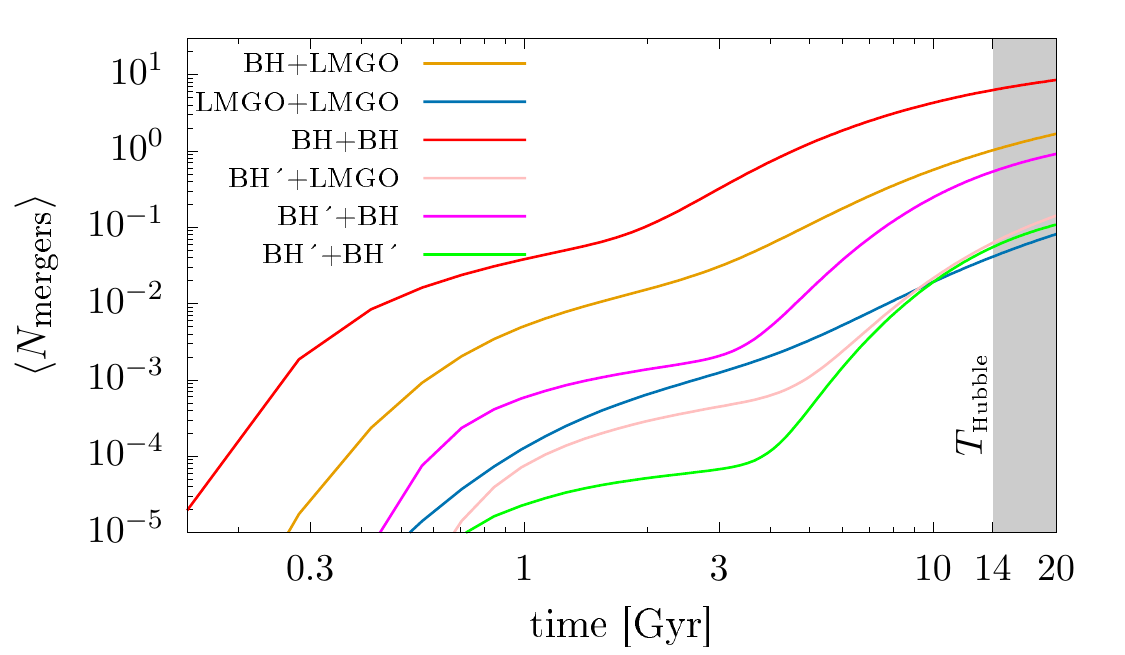}
	\includegraphics[width=0.5\textwidth]{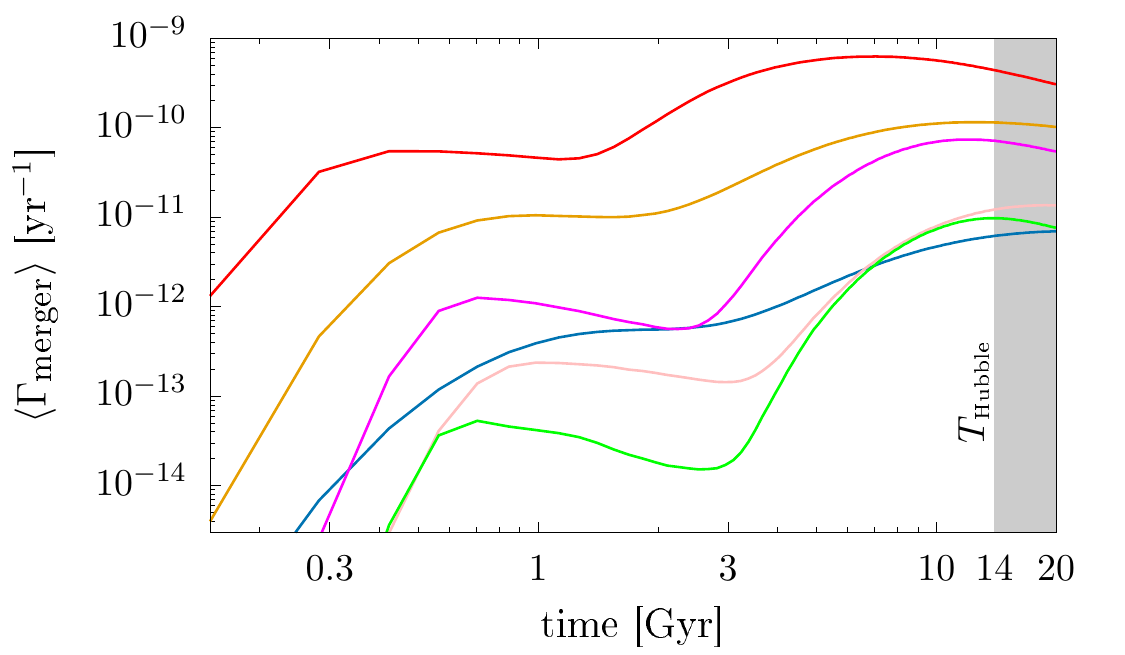}
	\caption{Globular clusters dominating in different epochs. There is a contribution coming from dense small clusters 
	which evolve rapidly as M30. Later a more dominant contribution comes from the larger clusters that 
	evolve slower. \textit{Top}: cumulative number per cluster, \textit{bottom}, averaged merger rate per cluster. We use the 
	same color coding in both panels.}
	\label{fig:GChistories}
\end{figure}

We point out that in the rates mentioned here, we took the averaged over 10 Gyr rates from the 106 Milky Way globular
clusters as being constant. That is an approximation that underestimates the rates at lower redshifts. Each cluster requires 
some time for the dynamical interactions to take place. Thus, there are very few mergers in the first Gyrs in most cluster's 
history, while those rates increase significantly at the last few Gyrs (see e.g. Fig.~\ref{fig:prototypical_All}). Each cluster 
due to its own density and mass has each own history of mergers. Small dense clusters evolve rapidly while more massive 
ones produce mergers later due to their smaller central densities (as 47Tuc). This is shown in Fig.~\ref{fig:GChistories} where 
we give the number of cumulative mergers and the equivalent rate per cluster.

While we can not directly assume the exact same evolution histories for globular clusters in other galaxies, 
Fig.~\ref{fig:GChistories} allows us to include the redshift evolution of the BH-LMGO mergers inside globular 
clusters (calculated in per cluster units). Assuming no build-up on the LMGO number we get a total merger rate density of 
$\Gamma_\text{BH-LMGO}\simeq (5-16)\times10^{-2} \text{Gpc}^{-3}\text{yr}^{-1}$. The main effect of that correction 
is the larger by a factor of two rate density. This is a direct result of the dominant contribution from massive globular clusters
at late times. 

\section{Conclusions and Discussion}
\label{secConclusions}

In this work we modeled the mechanism through which compact binaries, involving 
low-mass gap objects and/or black holes, form via exchange interactions inside 
Milky Way globular clusters.
The formed compact object binaries undergo hardening via interactions with stars in the
same environments leading to their accelerated merger.  Our basic scheme of interactions
 is depicted in Fig.~\ref{diagram}. 

For simplicity, we took monochromatic mass spectra for both the LMGOs, the first generation
black holes and the stars. We also included second generation black holes which result from 
merging first generation black hole binaries. Those second generation black holes can affect 
the dynamics inside the clusters and the merger rates of compact object binaries as those including
LMGOs. We took for the LMGOs number inside globular clusters to be directly related to the 
number of neutron stars that remain in those environments after their initial natal kick. 
We calculate the merger events for each type of binary using the observed Milky Way globular 
clusters from Ref.~\cite{harris1996}. We rely on their observed total mass and mass profile properties.
In Fig.~\ref{fig:prototypical_All}, we show for the Terzan 5 globular cluster the cumulative number 
of merger events and the evolution of the populations of individual compact objects and of binaries 
containing compact objects. 
From the Milky Way sample we then evaluated the averaged merger rates per cluster for each of 
the binary compact objects types, shown in Table~\ref{tab:values2}. 

We have found that the merger rate density of BH-LMGO mergers in the local Universe to be in the range 
$0.004-0.16\text{Gpc}^{-3}\text{yr}^{-1}$ up to a redshift of $z=1$. This result's range is associated mostly 
to how fast the LMGOs formed in the history of those clusters.  
This value is below the inferred rate from the GW190814 event i.e. 1-23
Gpc$^{-3}$yr$^{-1}$ but provides a lower estimate for these type of events. 
Thus globular clusters as set of environments can provide a population of such 
merger events. Depending on the individual cluster's mass profile and total mass, different 
types of clusters contribute at different epochs, with the more massive ones dominating the total
number of merger events at low redshifts (see Fig.~\ref{fig:2nd1stBBH}).
	
We do not evolve binaries involving stars, but rather monitor their population sizes.
Furthermore, we do not evolve the stellar cluster in time but instead integrate the evolution equations 
inside the volume in which the black holes uniformly populate, ignoring the possible existence of a 
massive BH at the center of the cluster. The spin of compact objects was not taken into account,
as analysis of the observed mergers suggests that black holes tend to have very small spins that are
narrowly distributed \cite{Miller:2020zox,Garcia-Bellido:2020pwq}. We have calculated the contribution
to the merger rates from direct capture events, i.e. hard compact object binaries forming from strong
dynamical encounters of individual objects and find it to be insignificant \footnote{For instance, 
in the early stages of Terzan 5, the capture merger rate of two first generation of BHs is 
$8.6\times10^{-13}\text{yr}^{-1}$. The corresponding rate for the exchange process ``B'' in 
Fig.~\ref{diagram} is $2.2\times10^{-8} \textrm{yr}^{-1}$ by comparison.}.

Three-body induced binaries involving compact objects are rare and typically form soft pairs 
that do not survive long enough inside the clusters. Thus we ignore them. We expect that 
binary-binary interactions can be of some importance in the evolution history of the clusters 
with the highest central densities (see e.g. \cite{Miller:2002pg, Antonini:2015zsa}). However, 
the total BH-LMGO, \BHprime-LMGO, BH-BH and \BHprime-BH merger rates are dominated 
by the most massive clusters, for which the binary-binary interactions only now (at $t \simeq 10$ Gyr) 
start having an impact. We leave that higher order correction in our calculations 
for future work, pointing out that it would only increase the claimed merger rates. 

The origin of the LMGOs is of importance. They represent a new category of objects with their 
own mass spectrum in the low mass region and could be of exotic origin as  primordial black holes 
formed in the early Universe. Here to be conservative we assumed that their totality is accounted 
for by the NS-NS mergers. Thus, our work provides a realistic lower estimate of the BH-LMGO
merger rate inside globular clusters.  We have focused on the impact this population has on 
the dynamics and formation of binaries involving at least one LMGO; leaving their origin as an
open question. 

\textit{Acknowledgements:} The authors are  grateful to A. Kehagias for his support in facilitating this collaboration. 
IC acknowledges the Faculty Research Fellowship support from the Oakland University Research Committee. 
	
\appendix
	
\section{Velocities and segregation radii}
\label{AppVelos}
	
According to the Virial theorem the three dimensional velocity dispersion of stars is given by 
$\sigma_*=\sqrt{{3\over5}{GM_\text{cl}\over r_h}}$, where $M_\text{cl}$ is the mass of the 
cluster and $r_h$ is the half-mass radius\footnote{The factor of 3/5 comes from 
integrating a uniform sphere of mass to obtain its potential and combine with scalar Virial theorem 
in steady state.}. 
In this context, stars represent the mean mass population, $\langle m\rangle=m_*$ as they 
are the most abundant objects in any globular cluster. Taking into account the fact that energy 
equipartition is not necessarily achieved in stellar clusters, \cite{Khalisi:2006ne,10.1093/mnras/stt1521}, 
the velocity of massive species with mass $m$ is given by $\sigma(m)=\left({m\over 2m_*}\right)^{-0.16}\sigma_*$ 
\cite{10.1093/mnras/stt1521}.
The relative velocity dispersion between two populations with masses $m_1$ and $m_2$ is given by,
\begin{equation}
\sigma_{1,2}=\sqrt{\sigma_1^2+\sigma_2^2}=\sqrt{m_1^{-0.32}+m_2^{-0.32}\over(2m_*)^{-0.32}}\sigma_*.
\end{equation}
	
Massive objects through dynamical friction sink towards 
the cluster's core partially segregating from the stars. Applying the Virial theorem for the segregated 
sub-population with objects of mass $m$, we find its half-mass radius $R_\text{seg}(m)$. This segregation 
radius is obtained by solving numerically the following algebraic equation, \cite{Kocsis:2006hq},
\begin{equation}
{4M(R_\text{seg}(m))\over M_\text{cl}}=\left({m\over2m_*}\right)^{-0.32}{R_\text{seg}(m)\over r_h}.
\end{equation}
$M(R_\text{seg}(m))$ is the cluster's mass up to $R_\text{seg}(m)$.
$M(R_\text{seg}(m))$ is simply evaluated for BHs that exist well within the core radius but is less trivial
for NSs or LMGOs that occupy a larger value and mass density of the cluster has a profile. We assume the 
King profile for the total mass density \cite{king}. 
As an example, the segregation volume of BHs with mass $m_\text{BH}$ is given by 
$V_\text{BH}={4\over3}\pi (R_\text{seg}(m_\text{BH}))^3$.

\section{The dependence of our results on input parameters}
\label{AppParameters}

In this work we used specific choices for the masses of the stars and compact objects 
to evaluate the merger rates of BH-LMGO and the other type of compact object binaries. 
In addition, the subsequent hardening rate for these compact object binaries from 3rd-body interactions with 
stars described in Eqs.~\ref{eq:SMAevolHard} and~\ref{eq:ECCevolHard}, is set by the choice
of the harding rates $H$ and $K$. In Table~\ref{tab:AltParameters}, we show how our main 
resulting per cluster averaged merger rates for BH-LMGO, \BHprime-LMGO and \BHprime-BH binaries 
change with alternative choices for each of these six parameters. 
We remind the reader that our reference choices in the main text were,  
$m_\text{BH}=10M_\odot$, $m_\text{LMGO}=3M_\odot$, $m_\text{star}=1M_\odot$, 
$H=20$ and $K = 0.05$. The mass of a second generation black holes is derived from
 the choice of first generation black hole to be $m_{\textrm{\BHprime}} = 1.9 \times m_\textrm{BH}$,
 i.e 19 $M_\odot$ in our reference case.
  
\begin{table}
\centering
\begin{tabular}{c c c c}
		\hline
		& BH+LMGO & \BHprime+LMGO & \BHprime+BH \\
		& & $\times10^{-2}$ & $\times10^{-1}$ \\
		\hline
		reference & 0.565 & 1.77 & 2.04 \\
		$m_\text{BH}=7M_\odot$ & 0.634 & 1.11 & 1.18 \\
		$m_\text{BH}=13M_\odot$ & 0.553 & 2.38 & 2.49 \\
		$m_\text{LMGO}=4M_\odot$ & 1.03 & 3.28 & 2.17 \\
		$m_\text{star}=0.3M_\odot$ & 0.619 & 1.93 & 1.61 \\
		$H=15$ & 0.482 & 1.25 & 1.21 \\
		$K=0$ & 0.473 & 1.07 & 1.21 \\
		$K=0.1$ & 0.770 & 2.99 & 3.19 \\
		\hline
\end{tabular}
\caption{The averaged cumulative number of BH-LMGO, \BHprime-LMGO and \BHprime-BH mergers 
per cluster at 10Gyr. The first row of data uses the reference values of $m_\text{BH}=10M_\odot$, 
$m_\text{LMGO}=3M_\odot$, $m_\text{star}=1M_\odot$, $H=20$ and $K=0.05$, as in Table~\ref{tab:values2}. 
In the following rows we vary the value of a single model variable, away from its reference one. }
\label{tab:AltParameters}
\end{table}

We find that most alternative choices to our reference ones, affect our merger numbers and 
subsequent rates by only up to $\simeq 50\%$, with the reference ones giving the 
representative values.  
For instance, changing the mass of stars to a smaller value of 0.3 $M_{\odot}$ and the impact 
it has on the \BHprime-BH mergers, decreasing it by a factor of $\simeq 20 \%$. This is simply 
the fact that the smaller mass stars remove less energy from the compact object binaries 
when interacting with them compared to the 1 $M_{\odot}$ ones, affecting in a less 
prominent manner the orbital evolution of these binaries (i.e. Eqs.~\ref{eq:SMAevolHard} 
and~\ref{eq:ECCevolHard}). Interestingly, the choices of alternative values for $H$ and $K$ 
matter less. 
The one choice of some importance to our results is changing the LMGOs mass from 3 
to 4 $M_{\odot}$. The larger mass choice increases the BH-LMGO and \BHprime-LMGO
merger rates by $\simeq 2$, as the more massive LMGOs have larger exchange 
cross-sections in channels ``A'', ``C'', ``J'', ``L'' and ``R'' of Fig.~\ref{diagram}. 

\begin{figure}
\centering
\includegraphics[width=0.5\textwidth]{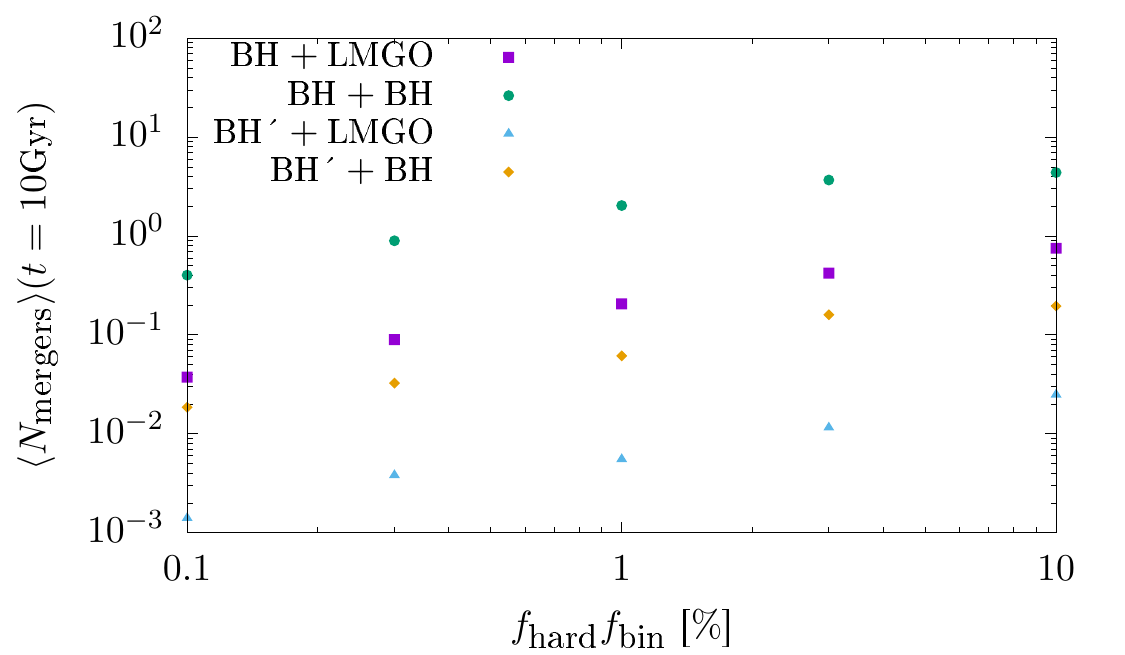}
\includegraphics[width=0.5\textwidth]{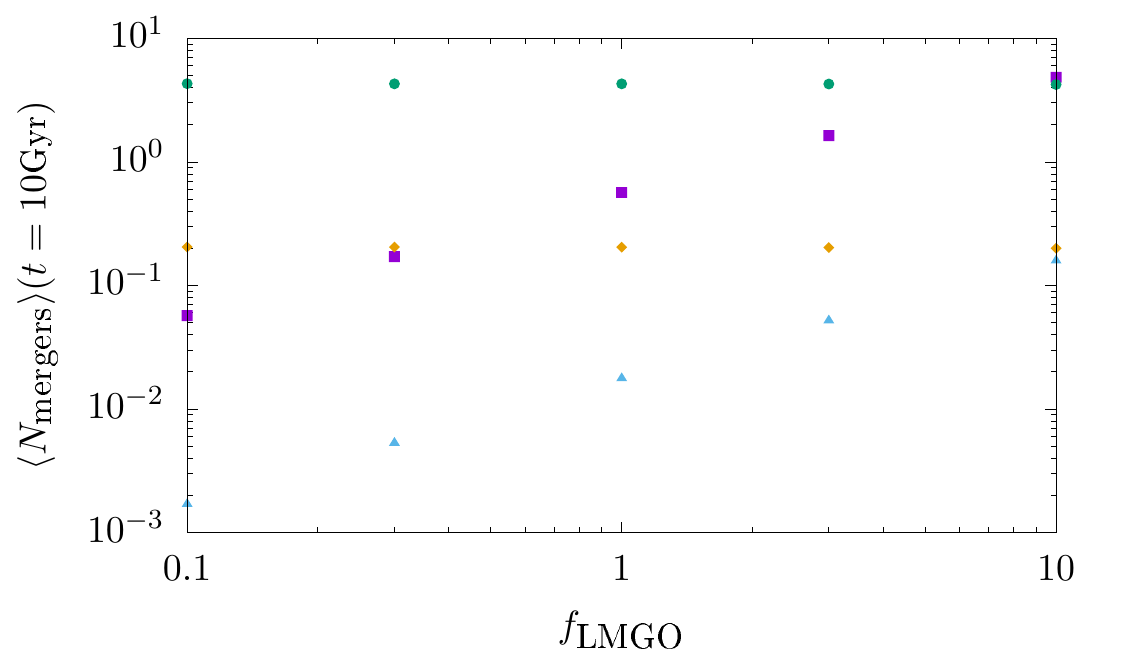}
\caption{The averaged number of cumulative mergers in Milky Way globular clusters, 
for BH-LMGO, \BHprime-LMGO, BH-BH and \BHprime-BH  binaries at 10Gyr. {\it Top panel}, 
we plot the number of mergers evaluated using various choices of initial hard star-star abundances.  
{\it Bottom panel}, for the same type of merging compact object binaries we change the initial 
abundance of single LMGOs, parameterized as $f_\text{LMGO}=2N_\text{LMGO}/N_\text{NS}$.
In the upper panel we took $f_\text{LMGO}=1$ (our reference choice) and in the lower panel 
$f_\text{hard}f_\text{bin}=0.05$.}
\label{fig:fHfB_plot}
\end{figure}

We have also assumed in the main text that the fraction of stars in hard binaries 
$f_{\textrm{hard}} f_{\textrm{bin}}$ is 0.05. In Fig.~\ref{fig:fHfB_plot}, we show (at the 
top panel) how varying this choice affects the number of BH-LMGO, \BHprime-LMGO, 
BH-BH and \BHprime-BH mergers after 10 Gyrs of evolution, using the entire set of
106 Milky Way globular clusters. The low order approximation in the dynamics we 
consider in this work (i.e. binary-single exchanges) holds for $f_\text{hard}f_\text{bin}$ 
up to about $0.05$. For higher abundances of star-star binaries, the number of single 
compact objects quickly depletes as they form binaries with stars. 
Since we only include binary-single interactions there are very few compact objects 
that are left to participate in further binary-single exchanges. If we had included
binary-binary interactions the number of cumulative mergers would only  
matter after 8 Gyrs of evolution. These binary-binary higher order corrections in the 
dynamics are insignificant in the first Gyr of the massive clusters' evolution, and become 
relevant only if the number of compact object-star binaries increases above the corresponding 
single compact object population which happens only around 9 Gyrs of evolution. 
For smaller clusters where the evolution is faster due to their higher central densities the
binary-binary exchanges are more important but their contribution to the total 
merger rate is minuscule as we show in Fig.~\ref{fig:GChistories}.

Finally, in Fig.~\ref{fig:fHfB_plot} in the lower panel we show the average BH-LMGO, 
\BHprime-LMGO, BH-BH and \BHprime-BH mergers after 10 Gyrs, for different choices on 
the abundance of LMGOs in globular clusters. We define $f_\text{LMGO}=2N_\text{LMGO}/N_\text{NS}$, 
where in the main text we took $f_\text{LMGO}=1$. Changing this number is relevant 
if there are additional sources of LMGOs, as primordial black holes in that mass range. The 
number of BH-LMGO and \BHprime-LMGO mergers is approximately proportional to $f_\text{LMGO}$ 
in the range shown. Instead, the numbers of merging binaries not involving a LMGO are 
roughly independent of $f_\text{LMGO}$.	

\section{Retaining second generation black holes in clusters}
\label{App:2ndGenBHsKicks}
	
In this appendix we describe how we evaluate the retention fraction of second generation black holes. 
In obtaining this fraction we make two assumptions. Our first assumption is that the 
distribution of binaries is uniform in mass ratio. The second is that the GW kick velocity 
follows a Maxwell-Boltzmann distribution. Our first assumption is a conservative one as 
binaries in a dense environments tend to exchange their members toward a mass ratio 
closer to unity.

The gravitational radiation from a BH binary composed of unequal masses is not 
axisymmetric. By conservation laws, the momentum carried away by the GW imparts 
a kick into the binary which recoils at the moment of merger where the GW radiation 
reaches its maximum amplitude, \cite{Hughes:2004ck,1983MNRAS.203.1049F}. For a binary $m_1$, 
$m_2\le m_1$ and mass ratio $q=m_2/m_1$, the magnitude of the GW recoil calculated for a circular 
orbit to 2nd post-Newtonian order and ignoring the spin of the BHs is given by \cite{maggiore,Lousto:2007db}, 
\begin{equation}
V_\text{gw}(q)\simeq12\text{Mm/s}{q^2(1-q)\over(1+q)^5}\left[1-0.93{q\over(1+q)^2}\right].
\label{Vgw_eq}
\end{equation}
This function in Eq.~\eqref{Vgw_eq} obtains its maximum value $175$km/s when the mass ratio is $0.362$.
As the kick velocity follows a Maxwell-Boltzmann 
distribution with parameter $V_\text{gw}(q)/\sqrt{3}$, the fraction of binary black holes 
retained in a cluster is,
\begin{eqnarray}
	f_\text{gw}(V_\text{esc},q)&=&\text{erf}\left(\sqrt{3\over2}{V_\text{esc}\over V_\text{gw}(q)}\right) \\
	&-&\sqrt{6\over\pi}{V_\text{esc}\over V_\text{gw}(q)}\exp\left[-{3\over2}\left({V_\text{esc}\over V_\text{gw}(q)}\right)^2\right]. \nonumber
\label{fgwEQ}
\end{eqnarray}

In order to estimate the fraction of second generation \BHprime  that remain in the cluster we assume that 
the mass ratio of BH-BH pairs is uniformly distributed between $q_\text{min}$ and 1. Taking a BH mass function, \cite{Kovetz:2016kpi,LIGOScientific:2018jsj}
\begin{align}
	P(m)\simeq16\times m^{-2.35}\ e^{-m/(40M_\odot)}\Theta(m-5M_\odot),
	\label{pmEQ}
\end{align}
we estimate this lower value by considering three possible combinations of the form 
$q_\text{min}={\langle m\rangle\over\langle m\rangle+\Delta m}$, 
$q_\text{min}={\langle m\rangle-\Delta m\over\langle m\rangle}$ and 
$q_\text{min}={\langle m\rangle-\Delta m\over\langle m\rangle+\Delta m}$. 
$\langle m\rangle\simeq10M_\odot$ is the first moment of the distribution in 
Eq.~\eqref{pmEQ} and $\Delta m\simeq4.5M_\odot$, giving us 
$q_\text{min} = \int_{\langle m\rangle-\Delta m}^{\langle m\rangle+\Delta m}dm\ P(m)\simeq 0.7$. 
The uniform character and hard cutoff at $q_\text{min}$ is a choice that results in 
small numbers of retained \BHprime s.  
As the majority of pairs are composed of nearly equal mass members, the mass 
ratio will be tilted towards unity. A more realistic mass-ratio distribution 
would yield higher retention fractions.

Integrating  Eq.~\eqref{fgwEQ} over $q$ gives us the total retention fraction, 
$F_\text{gw}(V_\text{esc})$, of second generation BH's. Our result is depicted 
in Eq.~\ref{retFracGW} of the main text.
	
\bibliography{literature}	
\end{document}